\documentclass[twocolumn,showpacs,preprintnumbers,amsmath,amssymb]{revtex4}
\usepackage[english]{babel} 
\usepackage[utf8]{inputenc}
\usepackage[T1]{fontenc}
\usepackage{lmodern}
\usepackage{dsfont}
\usepackage{amsfonts}
\usepackage{slashed}
\usepackage{graphicx}
\usepackage{color}
\usepackage{multirow}
\usepackage{hyperref}
\usepackage{enumitem}
\usepackage{epsfig,dsfont}

\newcommand{\I}{\mathrm{i}}

\begin{document}
\title{New DoS approaches to finite density lattice QCD}
\author{Christof Gattringer, Michael Mandl and Pascal T\"orek}
\vskip1mm
\affiliation{University of Graz, Institute for Physics,  A-8010 Graz, Austria\footnote{Member of NAWI Graz.}}
\begin{abstract}
We present two new suggestions for density of states (DoS) approaches to finite density lattice QCD. Both proposals are 
based on the recently developed and successfully tested DoS FFA technique, which is a DoS approach for bosonic 
systems with a complex action problem. The two different implementations of DoS FFA we suggest for QCD 
make use of different representations of finite density lattice QCD in terms of suitable pseudo-fermion path integrals. 
The first proposal is based on a pseudo-fermion representation of the grand canonical QCD partition sum, while the 
second is a formulation for the canonical ensemble. We work out the details of the two proposals and discuss the 
results of exploratory 2-d test studies for free fermions at finite density, where exact reference data allow one to verify 
the final results and intermediate steps. 
\end{abstract}
\maketitle

\section{Introduction}

Finding a suitable approach for Monte Carlo simulations of finite density QCD that extends the accessible part of 
the QCD phase diagram towards larger values of the chemical potential is currently one of the great challenges for
lattice field theory. The problem is that at finite chemical potential the fermion determinant is complex and cannot 
be used as a probability in a Monte Carlo process. Among the approaches that have been explored to solve this so-called
complex action problem are density of states (DoS) techniques, introduced to lattice field theory in 
\cite{Gocksch:1987nt,Gocksch:1988iz}. The key challenge for a DoS approach is to compute the density with 
sufficiently high accuracy, such that it can be integrated over with fluctuating integrands that appear when 
evaluating observables at finite density. A naive determination with, e.g., simple histogram techniques turned 
out to be useful only for very low densities \cite{Schmidt:2005ap,Fodor:2007vv,Ejiri:2007ga,Ejiri:2012ng}. 

Inspired by the Wang-Landau \cite{WangLandau} approach an interesting new development was presented 
by Langfeld, Lucini and Rago  
\cite{Langfeld:2012ah,Langfeld:2013xbf,Langfeld:2014nta,Langfeld:2015fua,Garron:2016noc,Garron:2017fta,Francesconi:2019nph}. 
The idea is to use a parameterization of the density as the exponential 
$\rho(x) = \exp(-L(x))$ of a piecewise linear and continuous function $L(x)$. Vacuum expectation values restricted to
the intervals where $L(x)$ is linear are used to determine $\rho(x)$ with very high precision. A variant of the 
Langfeld-Lucini-Rago method is the DoS Functional Fit Approach (FFA) 
\cite{FFA_1,FFA_2,FFA_3,Z3_FFA_1,Z3_FFA_2} 
and in recent years both techniques were used to obtain interesting results for several bosonic lattice field theories 
at finite density (see, e.g., the review \cite{Gattringer:2016kco}).

However, no modern DoS formulation for systems with fermions was presented so far, and thus no clear path towards 
precise DoS calculations for finite density QCD has been outlined yet. The challenge is to formulate the DoS approach 
such that it is compatible with conventional pseudo-fermion Monte Carlo techniques that may be applied to a real and 
positive fermion determinant. 

In this paper we discuss two proposals how to implement the DoS FFA for finite density lattice QCD. The first of the 
two is based on using a suitable pseudo-fermion representation of the QCD grand canonical partition sum. The 
imaginary part of the action is identified and the density is considered as a function of that imaginary part. DoS FFA 
is used to determine the corresponding density and observables are then obtained as integrals of the density.  

The second proposal implements DoS FFA in a canonical setting. The canonical partition functions at fixed net quark 
number are written as the Fourier moments with respect to imaginary chemical potential $\mu = i \theta/\beta$. 
Considering $\theta$ as an additional degree of freedom in the path integral allows one to implement the DoS FFA 
and compute the density as a function of $\theta$. Observables at fixed net quark number are then again obtained as
integrals of the density. 

In both formulations only Monte Carlo simulations without sign problem are needed, which furthermore can be 
implemented using well established techniques of standard lattice QCD simulations. We
work out the details of the two new approaches and present the results of small exploratory 2-d studies of the free case
where exact results can be used to assess the results and intermediate steps of the new proposals. 

\section{General formulation of the DoS FFA approach}

Before we can discuss our two new DoS approaches to finite density QCD we first need to discuss the details
of the DoS formulation we use, the {\it Functional Fit Approach}. The FFA 
\cite{FFA_1,FFA_2,FFA_3} is here presented for a general bosonic theory with a complex action problem 
and we will later show that with suitably chosen pseudo-fermion representations finite density lattice QCD can 
be brought into the general form introduced in this section.

\subsection{Partition sum and density of states}

The vacuum expectation values $\langle  {\cal  O } \rangle$ for some observable ${\cal  O }$ that 
we consider here can be written as bosonic path integrals,
\begin{eqnarray}
\langle  {\cal  O } \rangle & = & \frac{1}{Z}  \int \!\! D[\Phi] \,
e^{\, - \, S_R[\Phi] \, + \, i \, \alpha \, X[\Phi]}  \; {\cal  O } [\Phi] \; ,
\nonumber \\
Z \; & = & \int \!\! D[\Phi] \,
e^{\, - \, S_R[\Phi] \, + \, i \, \alpha \, X[\Phi]}   \; ,
\label{Zbosonicdef}
\end{eqnarray}
where $\Phi$ denotes an arbitrary set of general bosonic lattice fields that can be based on sites or links, and
$\int \! {D}[\Phi]$ is the corresponding product measure. We have already separated the exponent of the 
Boltzmann factor into two terms, the real part $S_R[\Phi]$ of the action and the imaginary part $\alpha X[\Phi]$. 
We have allowed for a real-valued coupling $\alpha \in \mathds{R}$ multiplying the imaginary part, which is useful 
in some of the applications we have in mind. $S_R[\Phi]$, $\alpha X[\Phi]$ and the observable ${\cal  O } [\Phi]$ 
are real-valued functionals of the lattice fields, and $S_R[\Phi]$ is assumed to be bounded from below. Obviously 
the  imaginary part $X[\Phi]$ gives rise to a complex action problem. 

For setting up the density of states approach we define the densities 
\begin{equation}
\rho^{^{({\cal J})}\!}(x) \; = \; \int \!\! D[\Phi] \,
e^{\, - \, S_R[\Phi]} \; {\cal  J } [\Phi] \;  \delta\Big(x-X[\Phi]\Big) \; ,
\label{densities}
\end{equation}
where ${\cal J}[\Phi]$ is an arbitrary real and positive functional of the fields. Below we will identify  ${\cal J}[\Phi]$ with some observable which in general can 
be decomposed into pieces that obey these requirements. Note that different choices of ${\cal J}[\Phi]$ result in 
different densities $\rho^{^{({\cal J})}\!}(x)$ and we use a superscript ${\cal J}$ to indicate which density we refer to. 

With the densities $\rho^{^{({\cal J})}\!}(x)$ vacuum expectation values $\langle {\cal O} \rangle$ of observables
$ {\cal  O}$ can be expressed as
\begin{eqnarray}
\langle  {\cal  O} \rangle & = & \frac{1}{Z} \int \! dx \,\rho^{^{( {\cal O})}\!}(x)\,
e^{\, i \, \alpha \, x} 
\; , \; \;
Z \; =  \int \! dx \,\rho^{^{(\mathds{1})}\!}(x)\,
e^{\, i \, \alpha \, x} \; , 
\nonumber \\
&& \langle {\cal F}(X) \rangle \; = \; \frac{1}{Z} \int \! dx \,\rho^{^{( \mathds{1})}\!}(x)\, {\cal F}(x) \,
e^{\, i \, \alpha \, x} \; ,
\label{obsdens}
\end{eqnarray}
where in the second line we have explicitly listed also the particularly simple case where the observable
is some function ${\cal F}$ of the imaginary part $X[\Phi]$.

The range of integration for the integrals $\int \! dx$ in (\ref{obsdens}) 
depends on the properties of the imaginary part $X[\Phi]$. 
If $X[\Phi]$ is bounded by some number $x_{max}$, so is the integration range. We will see that this is the case for the canonical DoS formulation 
discussed in Section 4. Furthermore, usually one can identify symmetries to show that the densities
$\rho^{^{( {\cal J})}\!}(x)$ are even or odd (depending on ${\cal J}$), such that the integration interval where we need 
to determine the densities is $[0,x_{max}]$. 

In case $X[\Phi]$ is unbounded the integration runs up to $x = \infty$, which is the case we will encounter 
in the direct DoS approach discussed in Section 3. Again symmetries can be 
used to show that $\rho^{^{( {\cal J})}}\!(x)$ is even (or odd) such that the actual integral is $\int_0^\infty \! dx$. 
Furthermore we will see that the densities $\rho^{^{( {\cal J})}}\!(x)$ quickly decrease with $x$ such that the range 
of integration can be truncated such that also in this second case we need to determine the densities in an 
interval $x \in [0,x_{max}]$.

Having defined the densities $\rho^{^{({\cal J})}\!}(x)$ and expressed observables as integrals over 
these densities we now have to address the problem of finding a suitable representation of the densities and how to 
determine the parameters used in the chosen representation.

\subsection{Parametrization of the density}

The densities $\rho^{^{({\cal J})}\!}(x)$ are functions of the parameter $x$ and we are interested in the densities in some 
finite interval $[0,x_{max}]$. For parameterizing the densities, we divide the interval $[0,x_{max}]$ into $N$ subintervals 
as follows:
\begin{equation}
[0,x_{max}] \; = \; \bigcup_{j=0}^{N-1}I_j\;,\quad\text{with}\quad 
I_j=[x_j,x_{j+1}]\;,
\label{intervals}
\end{equation}
where $x_0=0$ and $x_N= x_{max}$. Let $\Delta_j \equiv x_{j+1}-x_j$ denote the length 
of the interval $I_j$, such that
\begin{equation}
x_n \; = \; \sum_{j=0}^{n-1}\Delta_j \; .
\label{xn}
\end{equation}
The densities $\rho^{^{({\cal J})}\!}(x)$ are now parameterized in the form 
\begin{equation}
\rho^{^{({\cal J})}\!}(x) \; = \; \exp \left( - \, L^{^{({\cal J})}\!}(x) \right) \; ,
\label{eq:dens}
\end{equation}
where the $L^{^{({\cal J})}\!}(x)$ are continuous functions that are piecewise linear on the intervals $I_j$. Furthermore
we require $L^{^{({\cal J})}}\!(0) = 0$, such that the densities are normalized to $\rho^{^{({\cal J})}\!}(0) = 1$. 
For every interval $I_n$ we introduce a constant $a_n^{^{({\cal J})}}$ and a slope $k_n^{^{({\cal J})}}$ 
for the linear function, i.e., 
\begin{equation}
L^{^{({\cal J})}\!}(x) \; = \; a_n^{^{({\cal J})}} + \; k_n^{^{({\cal J} )}} \big(x - x_n \big)\; \; \mbox{for} \;
x \in I_n  \; .
\end{equation}
Using the fact that the functions $L^{^{({\cal J})}\!}(x)$ are required to be continuous and are normalized with 
$L^{^{({\cal J})}\!}(0) = 0$ we can completely determine the constants $a_n$ as functions of the slopes $k_n$. 
A simple calculation shows that $L^{^{({\cal J})}\!}(x)$ can be written in the following closed form,  
\begin{eqnarray}
L^{^{({\cal J})}\!}(x) & = & d_n^{^{\,({\cal J})}} \, + \, x  \, k_n^{^{({\cal J})}} 
\; \; \mbox{for} \; \; 
x \in I_n  
\label{piecewise} \\
\mbox{with} &&
d_n^{^{\,({\cal J})}} \, = \, \sum_{j=0}^{n-1} \left( k_j^{^{({\cal J})}}\!-k_n^{^{({\cal J})}} \right)\Delta_j   \; ,
\nonumber 
\end{eqnarray}
from which we obtain the explicit form of the density $\rho^{^{({\cal J})}\!}(x)$ in an interval $I_n$,
\begin{equation}
\rho^{^{({\cal J})}\!}(x) \,= \, A_n^{^{({\cal J})}} e^{ \, -  \, x \, k_n^{^{({\cal J})}} } \!\!, \;\; 
A_n^{^{({\cal J})}} \! \!= \; e^{\, -  \, d_n^{^{\,({\cal J})}} } \; \mbox{for} \;
x \in I_n  \, .
\label{rho_interval}
\end{equation}
Thus our parameterized density $\rho^{^{({\cal J})}\!}(x)$ 
depends only on the set of slopes $k_n^{^{{({\cal J})}}}$, one for each of the intervals 
$I_n$. We point out that the parametrization allows one to work with intervals $I_n$ 
that have different sizes $\Delta_n$. In particular in regions where the density 
$\rho^{^{({\cal J})}\!}(x)$ varies quickly one should use smaller 
intervals, while in regions of slow variation larger $\Delta_n$ can be used to reduce the computational cost. 
For a coarse scan of the density $\rho^{^{({\cal J})}\!}(x)$ with the goal of determining the regions of quick 
variation, one can do a first numerically cheaper determination with large $\Delta_n$ which subsequently is refined with 
finer intervals. These techniques are referred to as {\it preconditioning} and are discussed in detail in 
\cite{FFA_1,FFA_2,FFA_3}.

\subsection{Evaluation of the density parameters with FFA}
To determine the density, we need to compute the slopes $k_n^{^{({\cal J})}}$. For this purpose we introduce the 
restricted expectation values $\langle \, X \, \rangle_n^{^{({\cal J})}}(\lambda)$ which are defined as
\begin{eqnarray}
\langle \, X \, \rangle_n^{^{\!({\cal J})}\!}(\lambda) & \equiv &
\frac{1}{Z_n^{^{\,({\cal J})}\!}(\lambda) }
\int \!\! D[\Phi] \; e^{-S_R[\Phi] \, + \,  \lambda \, X[\Phi] } \qquad \quad \;
 \label{Xrest} \\
&& \qquad \qquad \qquad \times \quad
X[\Phi]  \; {\cal J}[\Phi] \; \Theta_n \! \left( X[\Phi]  \right) \; ,
\nonumber
\end{eqnarray}
with the corresponding restricted partition sums $Z_n^{^{\,({\cal J})}\!}(\lambda)$ given by 
\begin{equation}
Z_n^{^{\,({\cal J})}\!}(\lambda) \; \equiv \; 
\int \!\! D[\Phi] \; e^{-S_R[\Phi] \, + \, \lambda \, X[\Phi] } \; {\cal J}[\Phi] \; \Theta_n \! \left( X[\Phi]  \right) \; ,
\label{Zrest}
\end{equation}
where we have introduced the support functions
\begin{equation}
\Theta_n(x) \; = \; \left\{ 
\begin{array}{cc}
1 & \mbox{for} \;\;\; x \in I_n \; ,\\
0 & \mbox{for} \;\;\; x \notin I_n \; .
\end{array}
\right. 
\end{equation}
In the restricted expectation values $\langle \, X \, \rangle_n^{^{\!({\cal J})}\!}(\lambda)$ and the partition sums
$Z_n^{^{\,({\cal J})}\!}(\lambda)$ we have introduced a free real parameter $\lambda$ which couples to the 
imaginary part $X[\Phi]$ and enters in exponential form. Varying this parameter allows one to properly 
explore the $x$-dependence of the density in the whole interval $I_n$. The expectation values 
$\langle \, X \, \rangle_n^{^{\!({\cal J})}\!}(\lambda)$ are free of complex action problems and can be evaluated 
using Monte Carlo simulations. 

However, $\langle \, X \, \rangle_n^{^{\!({\cal J})}\!}(\lambda)$ and $Z_n^{^{\,({\cal J})}\!}(\lambda)$ can be computed also 
in closed form when using the parameterized density $\rho^{^{({\cal J})}\!}(x)$ in the form of Eq.~(\ref{rho_interval}). 
For the partition sums one obtains 
\begin{eqnarray}
Z_n^{^{\,({\cal J})}\!}(\lambda)  & = & \!\! \int\limits_{x_n}^{x_{n+1}} \!\!\!\! d x \,\rho^{^{({\cal J})}\!}(x) \, e^{\, \lambda \, x} 
\; = \; e^{- \, d_n^{^{\,({\cal J})}}} \!\! \int\limits_{x_n}^{x_{n+1}} \!\!\!\! d x \, e^{- \, x \, k_n^{^{({\cal J} )}} } \! e^{\, \lambda \, x}
\nonumber \\
& = & e^{- \, d_n^{^{\,({\cal J})}}} \,\frac{e^{\,x_n \big[ \lambda-k_n^{^{({\cal J})}} \big]}}{\lambda-k_n^{^{({\cal J})}}}
\,\Big(e^{\Delta_n\big[\lambda-k_n^{^{({\cal J})}}\big]}-1\Big) \, .
\end{eqnarray}
In the first step we have rewritten the restricted partition sum as the integral of the density $\rho^{^{({\cal J})}}\!(x)$ over
the interval $[x_n,x_{n+1}]$. In the second step the parameterized form (\ref{rho_interval}) was inserted for that particular 
interval, which gives rise to a simple integral of an exponential that can be evaluated in the closed form on the right-hand side.

Comparing (\ref{Xrest}) and (\ref{Zrest}) it is obvious that the restricted vacuum expectation value  
$\langle \, X \, \rangle_n^{^{\!({\cal J})}\!}(\lambda)$ can be computed as the derivative 
$\langle \, X \, \rangle_n^{^{\!({\cal J})}\!}(\lambda) \; = \; d \, \ln Z_n^{^{\,({\cal J})}\!}(\lambda) / d \lambda $, such that we
find the closed expression,
\begin{eqnarray}
\langle \, X \, \rangle_n^{^{\!({\cal J})}}(\lambda) & = & \frac{d \ln Z_n^{^{\,({\cal J})}\!}(\lambda)}{ d \, \lambda} 
 \\ 
& = & x_n+\frac{\Delta_n}{1-e^{-\Delta_n\big[\lambda-k_n^{^{({\cal J})}}\big]}}
-\frac{1}{\lambda-k_n^{^{({\cal J})}}}\;.
\nonumber 
\end{eqnarray}
After multiplicative and additive normalization we can express the result for
$\langle \, X \, \rangle_n^{^{\!({\cal J})}}\! (\lambda)$ (which in its normalized form we denote as 
$V_n^{^{\,({\cal J})}\!}(\lambda)$) in terms of a function $h(s)$,
\begin{equation}
V_n^{^{\,({\cal J})}\!}(\lambda) \, \equiv \, \frac{ \langle \, X \, \rangle_n^{^{\!({\cal J})}\!}(\lambda) - x_n}{\Delta_n}
- \frac{1}{2} \, = \, h\Big(\Delta_n\big[\lambda-k_n^{{({\cal J})}}\big]\Big) \; ,
\label{Vdef}
\end{equation}
where $h(s)$ is defined as
\begin{equation} 
h(s) \, \equiv \,  \frac{1}{1-e^{-s}}-\frac{1}{s}-\frac{1}{2}  \; ,
\label{hdef}
\end{equation}
and has the properties
\begin{equation}  
h(0)=0 \; ,  \; h^\prime(s)=1/12\; , \; \lim_{s\to\pm\infty}h(s)=\pm 1/2 \; .
\label{hdef}
\end{equation}

The strategy for determining the slope $k_n^{^{({\cal J})}}$ for an interval $I_n$ now is as follows: 
Using a standard Monte Carlo simulation without sign problem we compute the restricted vacuum 
expectation value $\langle \, X \, \rangle_n^{^{\!({\cal J})}\!}(\lambda)$ for several values of $\lambda$. 
After bringing the $\langle \, X \, \rangle_n^{^{\!({\cal J})}\!}(\lambda)$ into the normalized form 
$V_n^{^{\,({\cal J})}\!}(\lambda)$ defined in (\ref{Vdef}), the data for different $\lambda$ can be fit with 
the function $h\big(\Delta_n\big[\lambda-k_n^{{({\cal J})}}\big]\big)$, where the slope $k_n^{^{({\cal J})}}$ 
appears as the only fit parameter. From the set of $k_n^{^{({\cal J})}}$ the density $\rho^{^{({\cal J})}\!}(x)$ 
then is determined using (\ref{piecewise}) and (\ref{rho_interval}). Finally, vacuum expectation values of 
observables are computed via the integrals (\ref{obsdens}).

\section{Direct DoS approach for lattice QCD with a chemical potential} 

In this section we discuss the first of our two implementations of the new DoS approach to finite density QCD. Here we 
use a suitable pseudo-fermion representation of the grand canonical partition sum and separate the part with the 
complex action problem. For this factor we set up the DoS FFA formulation, discuss its properties and present results
of a first exploratory test in the free case.

\subsection{Grand canonical partition sum and pseudo-fermion representation}

We consider lattice QCD with $N_f$ mass-degenerate flavors of Wilson fermions. After integrating out the fermions the 
corresponding grand canonical partition sum with quark chemical potential $\mu$ is given by 
\begin{equation}
Z(\mu) \; = \; \int \!\!  D[U]\; e^{-S_g[U]} \; \det {\cal D}[U,\mu]^{\,N_f} \; .
\label{Z_mu}
\end{equation}
We consider the theory in $d = 2$ and $d = 4$ dimensions using lattices of size $V = N_s^{d-1} \times N_t$.
The SU(3)-valued gauge variables $U_\nu(x)$ live on the links $(x,\nu)$ of the lattice and obey periodic 
boundary conditions.  Their path-integral measure is the product of Haar measures 
$\int \! D[U] = \prod_{x,\nu} \int_{SU(3)} dU_\nu(x)$. $S_G[U]$ is the Wilson gauge action 
(we dropped the constant additive term), 
\begin{eqnarray}
&& S_g[U] \; = \; - \frac{\beta_g}{3} \;  P[U] \;  ,
\label{SGdef} \\
&& P[U] \; = \! \sum_{x, \nu < \rho} \!\!  \mbox{Re} \; \mbox{Tr} \;
U_\nu(x) \, U_\rho(x + \hat{\nu}) \, U_\nu(x+\hat{\rho})^\dagger \, U_\rho(x)^\dagger \! .
\nonumber 
\end{eqnarray}
$\beta_g$ is the inverse gauge coupling and $P[U]$ the sum over the real parts of the traced plaquettes. 

By ${\cal D}[U,\mu]$ we denote the Wilson Dirac operator with chemical potential $\mu$ in the background of a gauge 
field configuration $U$. We write the Dirac operator in the form,
\begin{equation}
{\cal D}[U,\mu] \; = \; \mathds{1} \, - \, \kappa \, {\cal H}[U,\mu] \; , \; \; 
{\cal H}[U,\mu] \; = \; \sum_{\nu = 1}^d {\cal H}_\nu[U,\mu] \; ,
\label{Diracdef}
\end{equation}
with the matrix elements of the hopping terms given by
\begin{eqnarray}
{\cal H}_\nu[U,\mu]_{x,y} & = &
[\mathds{1} - \gamma_\nu] \, e^{\, \mu \, \delta_{\nu,d}} \, U_{\nu}(x) \, \delta_{x+\hat{\nu},y}  
\label{Hopdef}
\\
& + &
[\mathds{1} + \gamma_\nu] \, e^{\, - \mu \, \delta_{\nu,d}} \, U_{\nu}(x-\hat{\nu})^\dagger \, \delta_{x-\hat{\nu},y}  \; .
\nonumber 
\end{eqnarray}
By $\gamma_\nu$ we denote the Euclidean $\gamma$-matrices in $d = 2$ or $d = 4$ dimensions, 
and $\kappa$ is the hopping parameter $\kappa = 1/(2d + 2m)$ with $m$ the bare quark mass. 
To be specific, we use a representation of the Euclidean $\gamma$-matrices where $\gamma_d$ is 
symmetric, which in $d = 4$ is, e.g., the chiral representation and in $d = 2$ the choice
$\gamma_1 = \sigma_2, \gamma_2 = \sigma_1$ with $\gamma_5 = \sigma_3$.
In (\ref{Hopdef}) we use matrix/vector notation for the 
$d$ Dirac indices of the $\gamma$-matrices and the 3 color indices of the link variables
$U_\nu(x)$. The chemical potential gives different weight 
for hopping in forward and backward temporal direction, i.e., the $\nu = d$ direction. The fermions obey periodic boundary 
conditions in the spatial direction(s) and anti-periodic boundary conditions in time, i.e., the terms in (\ref{Hopdef}) 
that connect sites with $x_d = N_t - 1$ and $x_d = 0$ have an additional minus sign. 

In order to introduce a pseudo-fermion representation that is suitable for the DoS FFA we write the fermion determinant as
\begin{eqnarray}
\hspace*{-4mm}
&& \det {\cal D}[U,\mu]  =  \frac{\det {\cal D}[U,\mu]^\dagger \  \det {\cal D}[U,\mu]}{\det {\cal D}[U,\mu]^\dagger}
\label{pseudintro} \\
\hspace*{-4mm}
&& =  \; \det \! \left( {\cal D}[U,\mu]^\dagger \, {\cal D}[U,\mu] \right)  C \!\! \int \! \!\!{D}[\Phi] \,
e^{\, - \, \Phi^\dagger {\cal D}[U,\mu]^\dagger \Phi}
\nonumber \\
\hspace*{-4mm}
&& =  \;  \det \! \left( {\cal D}[U,\mu]^\dagger \, {\cal D}[U,\mu] \right)  C \!\! \int \!\!\! {D}[\Phi] \, 
e^{ - \Phi^\dagger \! {\cal A}[U,\mu]  \Phi \,+ \, i  \Phi^\dagger \! {\cal B}[U,\mu] \Phi  } 
\nonumber \\
\hspace*{-4mm}
&& =  \; \det \! \left( {\cal D}[U,\mu]^\dagger \, {\cal D}[U,\mu] \right) C \!\! \int \!\!\!{D}[\Phi] \,
e^{\, - \, S_R[\Phi,U]  \; + \; i \,X[\Phi,U] } \; .
\nonumber 
\end{eqnarray}
In the second step we have written $1/\det {\cal D}[U,\mu]^\dagger$ as a bosonic integral over a complex-valued
scalar field $\Phi(x)$ with $3 d$ components for Dirac and color degrees of freedom, and
in the exponent we use vector/matrix notation for all indices. The constant $C$ is given by $C = (1/2\pi)^{3d V}$. 
In the third step we have organized the exponent into real and imaginary parts such that the 
pseudo-fermion integral matches the general form introduced in (\ref{Zbosonicdef}),
where here we set $\alpha = 1$. The corresponding real and 
imaginary parts are given by 
\begin{equation}
S_R[\Phi,U] \, = \,  \Phi^\dagger \!{\cal A}[U,\mu] \Phi \; , \; \; 
X[\Phi,U] \, = \,  \Phi^\dagger \! {\cal B}[U,\mu] \Phi \, ,
\label{SG_X}
\end{equation}
where we also write the gauge field $U$ as an argument in $S_R[\Phi,U]$ and $X[\Phi,U]$ since the real and the 
imaginary parts depend on $U$ via the kernels ${\cal A}[U,\mu]$ and ${\cal B}[U,\mu]$. These two matrices 
are defined as 
\begin{eqnarray}
{\cal A}[U,\mu]  & = & \frac{ {\cal D}[U,\mu]  \; + \; {\cal D}[U,\mu]^\dagger }{2} \; , 
\nonumber \\
{\cal B}[U,\mu]  & = & \frac{ {\cal D}[U,\mu]  \; - \; {\cal D}[U,\mu]^\dagger }{2\, i}  \; .
\label{ABdef}
\end{eqnarray}
A straightforward evaluation gives the matrix elements,
\begin{eqnarray}
\hspace*{-5mm}
{\cal A}[U,\mu]_{x,y} & = & \mathds{1} \delta_{x,y} \; - \; \kappa \sum_{\nu = 1}^d \Gamma_\nu(\mu) \Big(
U_{\nu}(x) \, \delta_{x+\hat{\nu},y} 
\nonumber \\ 
&&  \hspace{20mm} + \;  U_{\nu}(x-\hat{\nu})^\dagger \, \delta_{x-\hat{\nu},y}  \Big) \; ,
\nonumber \\
\hspace*{-5mm}
{\cal B}[U,\mu]_{x,y} & = & - i \, \kappa \sum_{\nu = 1}^d \Gamma_\nu(\mu) \, \gamma_\nu \Big(
U_{\nu}(x) \, \delta_{x+\hat{\nu},y}  
\nonumber \\ 
&&  \hspace{20mm}  - \;  U_{\nu}(x-\hat{\nu})^\dagger \, \delta_{x-\hat{\nu},y}  \Big) \; ,
\label{ABelements}
\end{eqnarray}
where 
\begin{equation}
\Gamma_\nu(\mu) \, = \, \mathds{1}  \cosh( \mu \,\delta_{\nu,d}) \, - \, \gamma_d \sinh( \mu \, \delta_{\nu,d} )
\, = \, e^{\, - \, \gamma_d \, \mu \, \delta_{\nu,d}  } \, .
\end{equation}
These explicit forms of ${\cal A}[U,\mu]$ and ${\cal B}[U,\mu]$ will be useful when discussing properties of the 
DoS FFA below. 

Obviously  ${\cal A}[U,\mu]$ and ${\cal B}[U,\mu]$ are hermitian, such that the two quadratic forms for 
$S_R[\Phi,U]$ and $X[\Phi,U]$ defined in (\ref{SG_X}) are real. Thus the pseudo-fermion integral in 
(\ref{pseudintro}) has the form that allows one to use DoS FFA to evaluate that integral. This will be discussed 
in more detail in the next section.

Let us add a few comments on the first factor in (\ref{pseudintro}), i.e., the determinant 
$\det \big( {\cal D}[U,\mu]^\dagger \, {\cal D}[U,\mu] \big)$. Using the well known generalized $\gamma_5$-hermiticity
property ${\cal D}[U,\mu]^\dagger = \gamma_5 {\cal D}[U,-\mu] \gamma_5$ we find 
\begin{equation}
\hspace*{-3mm}
\det \! \big( {\cal D}[U,\mu]^\dagger  {\cal D}[U,\mu] \big) =  \det \! \big( {\cal D}[U,\!-\mu] \big) \! 
\det \! \big( {\cal D}[U,\mu] \big),
\label{gamma5herm}
\end{equation}
which shows that $\det \big( {\cal D}[U,\mu]^\dagger \, {\cal D}[U,\mu] \big)$ corresponds to the fermion determinant
of two mass-degenerate quark flavors with an isospin chemical potential which is free of complex action problems. 
We stress, however, that this isospin determinant is of course only a part of the weight and its coupling to the 
pseudo-fermion factor in (\ref{pseudintro}) generates the full dynamics (see also the comments below).

The matrix ${\cal D}[U,\mu]^\dagger \, {\cal D}[U,\mu]$ is obviously hermitian and has real 
and non-negative spectrum, such that it is directly accessible with pseudo-fermion methods. Possible approaches 
are a direct pseudo-fermion representation (below 
$\chi$ and $\chi_j$ denote bosonic complex-valued pseudo-fermion fields),
\begin{equation}
\hspace*{-3mm}
\det \! \big( {\cal D}[U,\mu]^\dagger  {\cal D}[U,\mu] \big)  \propto \! \int \! \!\!D[\chi] \, 
e^{ -  \chi^\dagger ( {\cal D}[U,\mu]^\dagger  {\cal D}[U,\mu] )^{-1} \chi}  ,
\label{PF1}
\end{equation}
or an order-$n$ Chebychev multi-boson representation \cite{Cheb1,Cheb2} of the form 
\begin{eqnarray}
&& 
\det \! \big( {\cal D}[U,\mu]^\dagger \, {\cal D}[U,\mu] \big) 
\label{PF2} \\
&& \qquad \propto  
\prod_{j=1}^n \frac{1}{\det \! \big( u_j - \kappa {\cal H}[U,\mu]\big)^{\!\dagger}}  \;
\frac{1}{\det \! \big( u_j - \kappa {\cal H}[U,\mu]\big)^{\phantom{\dagger}}} 
\nonumber \\
&& \qquad \propto   
\prod_{j=1}^n \int \!\! D[\chi_j] \, 
e^{\, - \, \chi_j^\dagger \big( u_j - \kappa {\cal H}[U,\mu]\big)^{\!\dagger} \! 
\big( u_j - \kappa {\cal H}[U,\mu] \big) \chi_j}  \; ,
\nonumber
\end{eqnarray} 
where $u_j = e^{i 2 \pi j / (n+1)}$ are the coefficients for the Chebychev factorization and we have 
used (\ref{Diracdef}) to write the Dirac operator using the hopping matrix ${\cal H}[U,\mu]$. 

For both pseudo-fermion representations (\ref{PF1}) and (\ref{PF2}) a necessary condition is that the spectrum of
${\cal D}[U,\mu] = \mathds{1} - \kappa \, {\cal H}[U,\mu]$ does not touch the origin. Obviously a sufficient condition for this 
to hold is $\| {\cal H}[U,\mu] \| < \kappa^{-1} = 2d+2m$, where we use the matrix norm 
$\| M \| = \sup_{\{ \vec{v}: \|\vec{v}\|=1 \}} \sqrt{ {{\vec{v}}^{\, \dagger} M^\dagger M \vec{v} }}$.  
A simple crude estimate for the norm $\| {\cal H}[U,\mu] \|$
can be obtained as follows: Using the triangle inequality one finds 
\begin{eqnarray}
\| {\cal H}[U,\mu] \| & \leq & \sum_{\nu = 1}^d \| {\cal H}_\nu [U,\mu] \| 
 \\
& = & 
\sum_{\nu = 1}^d  \sup_{\{ \vec{v}: \|\vec{v}\|=1 \}} \sqrt{ {\vec{v}}^{\,\dagger}  {\cal H}_\nu [U,\mu]^\dagger \, 
{\cal H}_\nu [U,\mu] \vec{v}} \; .
\nonumber
\end{eqnarray}
Using the definition (\ref{Hopdef}) of the hopping matrices ${\cal H}_\nu [U,\mu]$ and the projector properties 
$[\mathds{1} \pm \gamma_\nu] [\mathds{1} \mp \gamma_\nu] = 0$ and $[\mathds{1} \pm \gamma_\nu]^2 \, = \,
2 [\mathds{1} \pm \gamma_\nu]$ one finds in a few lines of algebra
\begin{equation}
{\cal H}_\nu [U,\mu]^\dagger {\cal H}_\nu [U,\mu] = \left\{ \!\! 
\begin{array}{c}
4 \mathds{1} \; , \quad \nu = 1, \, ... \; d-1  \; , \\
4 \cosh(2\mu) \mathds{1} - 4 \sinh(2\mu) \gamma_d  \, , \;  \nu = d .
\end{array} \right.
\end{equation}
The matrix $4 \cosh(2\mu) \mathds{1} - 4 \sinh(2\mu) \gamma_d$ has eigenvalues $e^{2 \mu}$ and $e^{-2 \mu}$
(twice degenerate for $d = 4$) such that  $\| {\cal H}_\nu [U,\mu] \| = 2$ for $\nu = 1, \, ... \; d-1$ and 
$\| {\cal H}_d [U,\mu] \| = 2e^\mu$. Consequently $\| {\cal H}[U,\mu] \| \; \leq 2(d-1) + 2e^\mu$, and the sufficient 
condition for the spectrum of ${\cal D}[U,\mu] $ to not touch the origin reads, 
\begin{equation}
2(d-1) + 2e^\mu <  2d + 2m \, \Leftrightarrow \, \mu <  \ln ( 1 + m)  = m  +  O(m^2) \, .
\label{muestimate}
\end{equation}  
Thus we find that there is a finite range of $\mu$ where the factor 
$\det \! \big( {\cal D}[U,\mu]^\dagger \, {\cal D}[U,\mu] \big)$ which is free of the complex action problem can be treated
with conventional pseudo-fermion techniques. We stress again, that the estimate (\ref{muestimate}) is only a crude 
non-exhaustive bound that essentially reflects the situation of the free case, where condensation sets in 
at $\mu = m$. For a dynamical background gauge configuration $U$ the spectrum of ${\cal H}[U,\mu]$ is
known to contract such that values of $\mu$ that exceed the bare quark mass parameter $m$ become accessible. 
To precisely delimit the range where the pseudo-fermion treatment of 
$\det \! \big( {\cal D}[U,\mu]^\dagger \, {\cal D}[U,\mu] \big)$ is possible beyond the bound (\ref{muestimate}) 
obviously is a dynamical question that has to take into account the emerging finite density physics as well as
possible numerical instabilities of the HMC algorithm that can only be assessed in a full QCD simulation, 
which clearly goes beyond the scope of this presentation. However, already with the simple bound (\ref{muestimate}) 
we have established an interesting minimal region where the direct DoS approach is applicable in principle.

\subsection{Implementation of the DoS FFA}

To set up the density of states approach and to define the corresponding densities as outlined in the general presentation 
in Section 2 we need to write the grand canonical definition with the pseudo-fermion representation. Since we consider 
the general case of $N_f$ flavors, we need $N_f$ copies of the pseudo-fermion fields, $\Phi_j, j = 1, \, ... \, N_f$,
where by $\{ \Phi \}$ we denote the set of all these fields. Based on the discussion of the previous section we thus 
write the grand canonical partition sum of QCD in the form that matches Eq.~(\ref{Zbosonicdef}) with 
$\alpha = 1$ (irrelevant overall constants were dropped), i.e.,
\begin{equation}
Z(\mu)  =  \! \int \!\!\!D[U]\, e^{\,-\,S_{eff}[U]} \! \int \! \! \! D[\{\Phi\}] \,
e^{\, - \, S_R[\{\Phi\},U]  \, + \, i \,X[\{\Phi\},U] } ,
\end{equation}
where the real and imaginary parts of the pseudo-fermion action, as well as the path integral measure were 
generalized to $N_f$ flavors,
\begin{eqnarray}
S_R[\{\Phi\},U]  & = & \sum_{j=1}^{N_f}  S_R[\Phi_j,U] \; = \; \sum_{j=1}^{N_f} \Phi_j^\dagger \, {\cal A}[U,\mu] \, \Phi_j \; ,
\nonumber \\
X[\{\Phi\},U]  & = & \sum_{j=1}^{N_f}  X[\Phi_j,U] \; = \; \sum_{j=1}^{N_f} \Phi_j^\dagger \, {\cal B}[U,\mu] \, \Phi_j \; ,
\nonumber \\
\int \! \!{D}[\{\Phi\}] & = & \prod_{j=1}^{N_f} \int \! \!{D}[\Phi_j] \; .
\label{Nf_defs}
\end{eqnarray}
As we have outlined in the previous
section the term $\det \! \big( {\cal D}[U,\mu]^\dagger \, {\cal D}[U,\mu] \big)$ in (\ref{pseudintro}) can be treated with
conventional pseudo-fermion techniques and we combined the corresponding factor for $N_f$ flavors together with 
the gauge field action $S_g[U]$ into the effective action $S_{eff}[U]$, 
\begin{equation}
e^{-S_{eff}[U]} \; = \; e^{-S_{g}[U]} \, \det \big( {\cal D}[U,\mu]^\dagger \, {\cal D}[U,\mu] \big)^{N_f} \; .
\label{Seff}
\end{equation}
Following the general DoS FFA strategy outlined in Section 2.1 we now define the densities as
\begin{eqnarray}
\hspace*{-10mm} 
&& \rho{^{^{({\cal J})}}\!}(x)  = \! \int \!\!\! D[U] \!\int \!\!\!{D}[\{\Phi\}]  \, e^{\, - \, S_{eff}[U] \, - \, S_R[\{\Phi\},U]} 
\label{densities_direct} \\
\hspace*{-10mm} 
&& \hspace{30mm} \times \; {\cal J}[\{\Phi\},U]  \; \delta \Big( x  -  \,X[\{\Phi\},U] \Big) \; ,
\nonumber
\end{eqnarray}
where we allow for general observables ${\cal J}[\{\Phi\},U]$ that can be functionals of both, the set $\{\Phi\}$
of pseudo-fermion fields $\Phi_j$, as well as the gauge fields $U$. 

In the general outline of the method in Section 2 we have already announced that symmetries 
can be used to establish that the densities  $\rho^{^{({\cal J})}\!}(x)$ are either even or odd functions, depending on the
observables ${\cal J}$, which we assume themselves to be even or odd (general ${\cal J}$ may be decomposed 
into even and odd pieces). As an example we briefly discuss the simplest case of ${\cal J} = \mathds{1}$ and show
that $\rho^{^{(\mathds{1})}\!}(x)$ is even. The symmetry transformation we consider is charge conjugation that for the 
gauge links and the pseudo-fermion fields is implemented as
\begin{eqnarray}
&& U_\nu(x) \; \rightarrow \; U_\nu(x)^\prime \; = \; U_\nu(x)^* \; \equiv \; \Big(U_\nu(x)^\dagger\Big)^T, 
\nonumber \\
&& \Phi_j(x) \; \rightarrow \; \Phi_j(x)^\prime \; = \; \Phi_j(x)^* \; \equiv \; \Big(\Phi_j(x)^\dagger\Big)^T ,
\label{trafo}
\end{eqnarray}
where $^*$ denotes complex conjugation and $^T$ transposition. It is straightforward to show that
\begin{eqnarray}
&& S_R[\Phi_j^\prime,U^\prime] \; = \; {\Phi_j^\prime}^\dagger {\cal A}[U^\prime\!,\mu] \Phi_j^\prime \; = \; 
S_R[\Phi_j,U] \; ,
\nonumber \\
&& X[\Phi_j^\prime,U^\prime] \; = \; {\Phi_j^\prime}^\dagger {\cal B}[U^\prime\!,\mu] \Phi_j^\prime \; = \; - \, X[\Phi_j,U] \; .
\end{eqnarray}
Equally straightforward is to show that the gauge action $S_g[U]$ defined in (\ref{SGdef}) is invariant under the 
charge conjugation transformation (\ref{trafo}), i.e., $S_g[U^\prime] = S_g[U]$. 

The invariance of the factor $\det \! \big( {\cal D}[U,\mu]^\dagger \, {\cal D}[U,\mu] \big)$ can be shown using 
the representation (\ref{gamma5herm}) and charge conjugation:
Denote by $C$ the charge conjugation matrix that obeys $C^{-1} \gamma_\nu C = - \gamma_\nu^{\;T}$. Then
\begin{eqnarray}
\hspace*{-5mm}&& \det \! \big( {\cal D}[U^\prime\!,\mu]^\dagger \, {\cal D}[U^\prime\!,\mu] \big) \, = \, 
\det\! \big( {\cal D}[U^\prime\!,-\mu] \big)  \det \big( {\cal D}[U^\prime\!,\mu] \big) 
\nonumber \\
\hspace*{-5mm}&&
 = \, \det \! \big(C^{-1} {\cal D}[U^\prime\!,-\mu] C\big)  \det \! \big( C^{-1} {\cal D}[U^\prime\!,\mu] C \big) 
 \\
\hspace*{-5mm}&&
= \, \det \! \big({\cal D}[U,\mu]^T \big) \!  \det \! \big( {\cal D}[U,-\mu]^T \big) 
\, = \, \det \! \big( {\cal D}[U,\mu]^\dagger {\cal D}[U,\mu] \big) ,
\nonumber
\end{eqnarray}
where in the first step we used (\ref{gamma5herm}), then inserted the charge conjugation matrix $C$ and finally 
exploited the relation $C^{-1} {\cal D}[U^\prime\!,\mu] C = {\cal D}[U,-\mu]^T$ which is easy to show for
the explicit form (\ref{Diracdef}), (\ref{Hopdef}) of ${\cal D}[U,\mu]$.

Finally, using the invariance of the path integral measures $\int \! D[U^\prime] \!=\! \int \! D[U]$ and 
$\int \! D[\{\Phi^\prime\}] \!= \! \int \! D[\{\Phi\}]$ we conclude
\begin{eqnarray}
\hspace*{-5mm} && \rho^{^{(\mathds{1})}\!}(x)  =  \int \!\!  D[U^\prime] \! \int \! \! {D}[\{\Phi^\prime\}]  \, 
e^{-S_{eff}[U^\prime] - S_R[\{\Phi^\prime\},U^\prime]} 
\nonumber \\
\hspace*{-5mm} && \hspace{30mm}
\times \; \delta \Big( x  - X[\{\Phi^\prime\},U^\prime] \! \Big)
 \\
\hspace*{-5mm}  && \hspace{10mm} = \!
\int \! \! D[U]  \! \int \!\!{D}[\{\Phi\}] \, e^{-S_{eff}[U] -  S_R[\{\Phi\},U]}  
\nonumber \\
\hspace*{-5mm} && \hspace{30mm}
\times \;
\delta \Big(  x  + X[\{\Phi\},U] \Big) = \rho{^{(\mathds{1})\!}}(-x) .
\nonumber
\end{eqnarray}
Thus we have established that $\rho^{^{(\mathds{1})}\!}(x)$ is an even function and in a similar way one may
analyze the symmetry properties for the general densities $\rho^{^{({\cal J})}\!}(x)$ that contain the insertion of
some observable ${\cal J}$.

As the final step for the implementation of the DoS FFA we need to identify the restricted expectation 
values as defined in the general description of the method in Section 2. Comparing with the general form 
(\ref{densities}), (\ref{Xrest}) we can read off from the densities (\ref{densities_direct}) the necessary restricted 
expectation values for full QCD,
\begin{eqnarray}
\hspace*{-4mm} && \langle  X \rangle_n^{^{\!({\cal J})}\!}(\lambda) 
\label{rest_vev_direct}  \\  
\hspace*{-4mm} && =  \frac{1}{Z_n^{^{({\cal J})}}\!(\lambda) } \!
\int \!\! D[U] \! \int \!\! D[\{\!\Phi\!\}] \, 
e^{\, - \, S_{eff}[U] \, - \, S_R[\{\!\Phi\!\},U] \, + \, \lambda \,X[\{\!\Phi\!\},U] } 
\nonumber \\ 
\hspace*{-4mm} && \hspace{25mm} \times \; \; X[\{\!\Phi\!\},U] \, {\cal J}[\{\!\Phi\!\},U] 
\, \Theta_n(X[\{\!\Phi\!\},U])  
\nonumber \\
\hspace*{-4mm} && =     
\frac{1}{Z_n^{^{({\cal J})}}\!(\lambda) } \!
\int \!\! D[U] \!  \int \!\! D[\{\!\Phi\!\}] \,  e^{-S_{eff}[U] \, - \sum_{j} \Phi_j^\dagger {\cal M}[U,\lambda] \Phi_j } 
\nonumber \\ 
\hspace*{-4mm} && \hspace{25mm} \times \; \; 
X[\{\!\Phi\!\},U]  \, {\cal J}[\{\!\Phi\!\},U] \, \Theta_n(X[\{\!\Phi\!\},U]) \; ,
\nonumber 
\end{eqnarray}
where in the second step we have written the combination $S_R[\{\!\Phi\!\},U]  -  \lambda \,X[\{\!\Phi\!\},U]$ as a 
sum of quadratic forms  
$S_R[\{\!\Phi\!\},U]   -  \lambda X[\{\!\Phi\!\},U] = \! \sum_{j} \Phi_j^\dagger {\cal M}[U,\lambda] \Phi_j$ 
with kernel
\begin{eqnarray}
{\cal M}[U,\lambda]_{x,y} & = &  \mathds{1} \delta_{x,y}  -  \kappa \! \sum_{\nu = 1}^d \! \Gamma_\nu(\mu)\Big( 
\! [\mathds{1} - i\lambda \gamma_\nu] U_\nu(x) \delta_{x+\hat{\nu},y}
\nonumber \\
&& + \; 
[\mathds{1} + i\lambda \gamma_\nu] \, U_\nu(x-\hat{\nu})^\dagger \, \delta_{x-\hat{\nu},y} \Big) \; .
\label{Mkernel}
\end{eqnarray}
Note that ${\cal M}[U,\lambda]$ is a hermitian matrix and for sufficiently small $\lambda$ its eigenvalues are 
positive (see the discussion below). Thus the restricted expectation values do not have a 
sign problem and can be computed with Monte Carlo simulations. After suitable 
normalization to the form (\ref{Vdef}) the corresponding functions $V_n^{^{\,({\cal J})}\!}(\lambda)$ can then be fit 
with the function $h(\Delta_j[\lambda - k_j^{^{({\cal J})}}])$. The results are the slopes 
$k_j^{^{({\cal J})}}$ which via (\ref{piecewise}) and (\ref{rho_interval}) determine the densities 
$\rho^{^{({\cal J})}\!}(x)$. Observables are then computed according to (\ref{obsdens}).  

We stress that the ensemble considered in the restricted vacuum expectation values (\ref{rest_vev_direct}) is not simply 
QCD with isospin chemical potential reweighted to quark chemical potential, where a serious overlap problem would 
emerge. Instead the exponent of the Boltzmann factor in (\ref{rest_vev_direct}) is given by 
$S_{eff}[U] \, + \, S_R[\{\!\Phi\!\},U] \, - \, \lambda \,X[\{\!\Phi\!\},U]$ where the isospin 
contribution that is hidden in $S_{eff}[U]$ is augmented with the contributions 
$S_R[\{\!\Phi\!\},U] \, - \, \lambda \,X[\{\!\Phi\!\},U]$ of the pseudo-fermion terms, which contribute the 
dynamics of the quark chemical potential.

\subsection{First tests for the free case}

For a first exploratory study of the new DoS approach we analyze the free case in two dimensions. 
Note that due to the restricted expectation
values that need to be evaluated, this analysis already requires Monte Carlo simulations also for the free case and 
indeed provides a non-trivial test of the method. Insight about suitable sizes $\Delta_n$ for the intervals, 
the numerical cost, the accuracy that is needed for the density et cetera, can be obtained. Furthermore, the free case 
allows for a systematical comparison of the final results and the intermediate steps against analytical results that may 
be computed with Fourier transformation.

For the free case the density $\rho^{^{(\mathds{1})}\!}(x)$ defined in (\ref{densities_direct}) with the help of the 
pseudo-fermion representation simplifies to (we consider the case of $N_f = 1$ flavor)
\begin{equation}
\rho^{^{(\mathds{1})}\!}(x)  \; = \; \int \!\!{D}[\Phi] \; 
e^{\, - \, S_R[\Phi]} \; \delta \Big( x  -  \,X[\Phi] \Big) \; . 
\label{densities_free}
\end{equation}
The gauge field integration has been dropped for the free case and also the Boltzmann factor (\ref{Seff}) for the 
effective action since it is independent of $x$, such that it only would affect the overall normalization of the density
which is set by requiring $\rho^{^{(\mathds{1})}\!}(0) = 1$. 

Following the steps of the implementation of 
DoS FFA in the previous section, for determining the parameters $k_j^{{(\mathds{1})}}$ we need to evaluate the 
restricted expectation values defined in (\ref{rest_vev_direct}) which for the free case reduce to 
\begin{equation}
\langle  X \rangle_n^{^{\!(\mathds{1})}\!}(\lambda) \, = \,    
\frac{1}{Z_n^{^{(\mathds{1})}}\!(\lambda) } \!
\int \!\! D[\Phi]  \,
e^{\, - \, \Phi^\dagger {\cal M}[\lambda] \, \Phi } \, X[\Phi]  \; 
\Theta_n(X[\Phi]) .
\label{rest_vev_free}
\end{equation}
The imaginary part $X[\Phi]$ can be obtained from (\ref{SG_X}) and (\ref{ABelements}) 
(drop the link variables $U_{\nu}(x)$ there) as,
\begin{equation}
X[\Phi] \, = \, - \, i \, \kappa \! \sum_{\nu = 1}^d \sum_x \Phi(x)^\dagger \, \Gamma_\nu(\mu) \, \gamma_\nu \Big(
\Phi(x+\hat{\nu})  \, - \,  \Phi(x-\hat{\nu}) \Big) ,
\end{equation}
and the kernel ${\cal M}[\lambda]$ in the Boltzmann factor of (\ref{rest_vev_free}) is given by (see (\ref{Mkernel})),
\begin{eqnarray}
{\cal M}[\lambda]_{x,y} & = & \mathds{1} \delta_{x,y} \; - \; \kappa \sum_{\nu = 1}^d \Gamma_\nu(\mu)\Big( 
[\mathds{1} - i\lambda \gamma_\nu] \,  \delta_{x+\hat{\nu},y} 
\nonumber \\
&& \qquad \qquad \qquad + \; 
[\mathds{1} + i\lambda \gamma_\nu] \,  \delta_{x-\hat{\nu},y} \Big) \; .
\label{Mkernel_free}
\end{eqnarray}
This matrix is obviously hermitian such that its eigenvalues are real. However, for the existence of the path integral 
needed for the evaluation of the restricted vacuum expectation values (\ref{rest_vev_free}), the eigenvalues also have to 
be positive, and we now address this issue that has been neglected previously. It is easy to see that eigenvalues
can become negative for large values of the parameter $\lambda$. For the application of DoS FFA we thus
need to establish that in the interesting region of the couplings $m$ and $\mu$ there is indeed a finite range of 
values of $\lambda$ such that in this range all eigenvalues are positive. 

The eigenvalues of ${\cal M}[\mu,\lambda]$ can easily be computed using Fourier transformation. From this exact 
result one finds that the spectrum is invariant under the reflection $\lambda \rightarrow - \lambda$. Thus for given 
couplings $m$ and $\mu$ the range of values $\lambda$ where all eigenvalues of  ${\cal M}[\mu,\lambda]$ are positive, 
must be a symmetrical interval $(-\lambda_{max}, +\lambda_{max})$. It is straightforward to determine this interval by 
analyzing the $\lambda$-dependence of the spectrum obtained from Fourier transformation. 

\begin{figure}[t]
\centering
\hspace*{-2mm}
\includegraphics[width=85mm,clip]{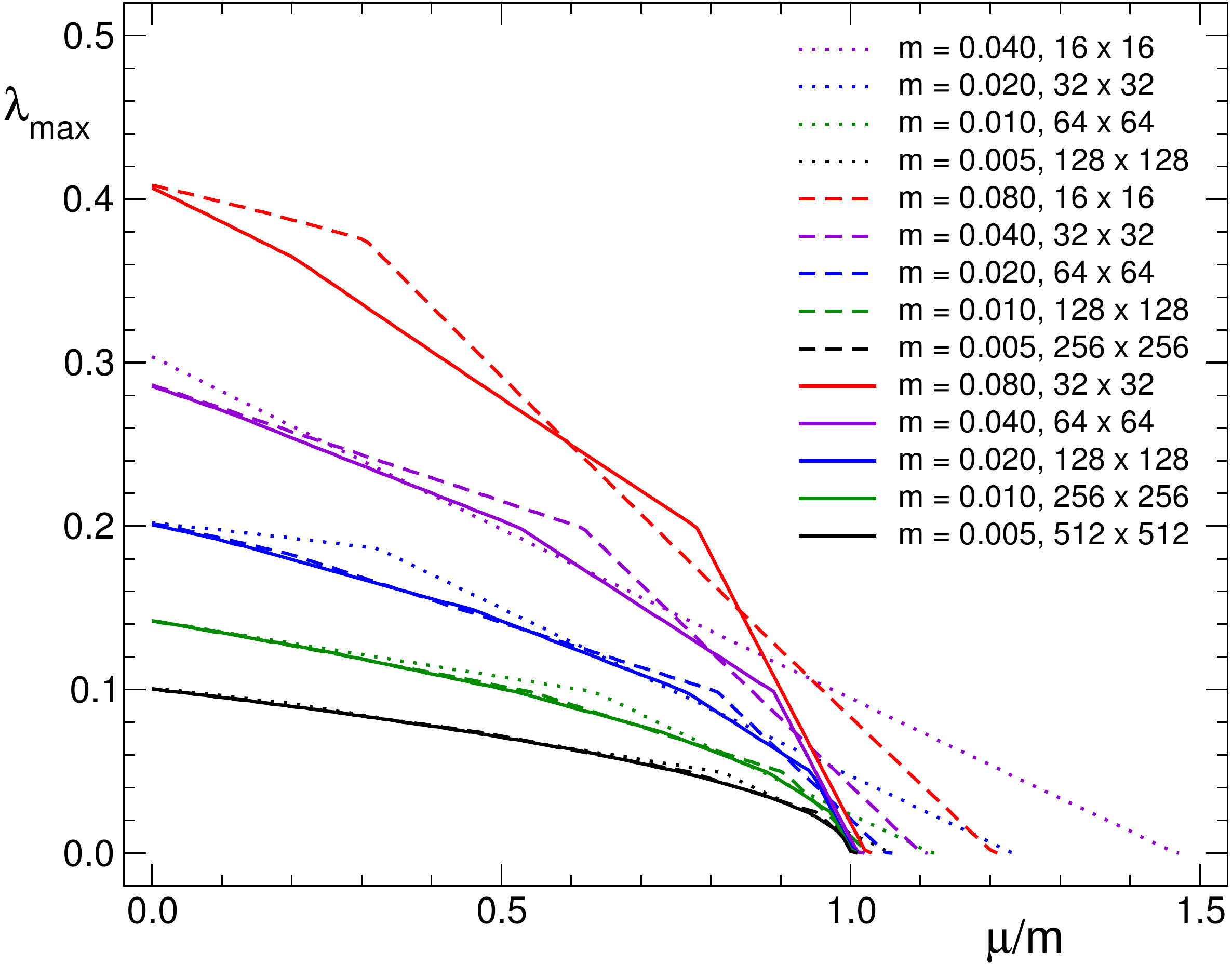}
\caption{The value $\lambda_{max}$ as a function of $\mu/m$. We compare the results for different values of the 
mass $m$ and different lattice sizes $L \times L$. The dotted curves correspond to fixed $m L = 0.64$, the dashed 
curves to $m L = 1.28$ and the full curves to $m L = 2.56$.} 
\label{lambda_max_plot}
\end{figure}

The result for such an analysis is shown in Fig.~\ref{lambda_max_plot}, where we plot the value $\lambda_{max}$ 
as a function of $\mu/m$ and compare our results for different masses $m$ and lattice sizes $L \times L$. 
The value of $\mu/m$ where $\lambda_{max}$ becomes zero signals 
the breakdown of the method. We remark that the polygon-like behavior of the curves for the smaller volumes reflects 
the fact that for small volumes the momenta populate the interval $[-\pi, \pi]$ with only a few values such that also
a ''sparse'' spectrum emerges, and the different sections of the ''polygon'' correspond to a different eigenvalue 
becoming negative. 

The data in Fig.~\ref{lambda_max_plot} is organized in groups where in each group we consider a sequence 
of values $L \rightarrow \infty$ and $m \rightarrow 0$ at a fixed value of $m L$, i.e., we study the fixed volume 
continuum limit of the free theory. The dotted curves are for $m L = 0.64$, the dashed curves for 
$m L = 1.28$ and the full curves correspond to $m L = 2.56$. Note that the curves for different $mL$ cluster 
according to the respective values of $m$. The figure shows that with increasing $\mu/m$ the values for 
$\lambda_{max}$ decrease and at a critical value of $\mu/m$ the boundary $\lambda_{max}$ becomes zero, 
signaling the breakdown of the method. We observe that for all three values of $mL$ we study, the critical
value of $\mu/m$ converges from above to a critical value of $\mu/m = 1$, which is the value of the chemical potential
where condensation sets in. Thus we expect that we can use DoS FFA all the way to the condensation point.

For the dynamical case one expects a similar behavior: For non-trivial gauge links the spectrum of the Dirac operator
is known to contract, giving rise to an additive renormalization of the mass and a critical $\kappa$ that is larger than the 
free value $\kappa = 1/2d$. Qualitatively one finds that for a larger critical $\kappa$ a larger value of $\mu$ is 
accessible, and one may expect that also for the full case the critical value of $\mu$ coincides with the point where 
condensation sets in. We stress, however, that obviously this is only a very qualitative discussion of the situation in the
fully dynamical case and future explicit Monte Carlo calculations will be necessary for a detailed analysis.

\begin{figure}[t]
\centering
\hspace*{-2mm}
\includegraphics[width=85mm,clip]{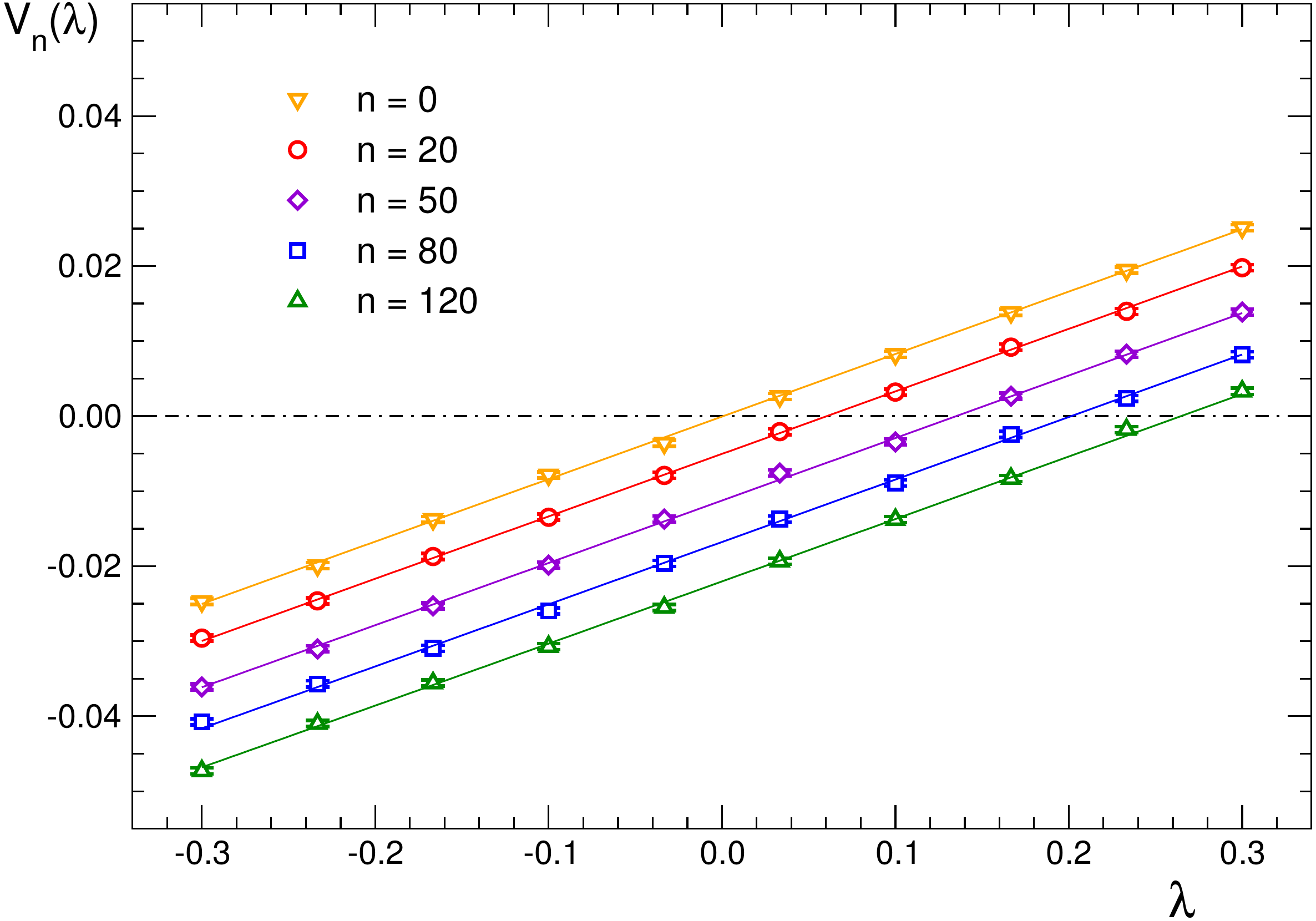}
\caption{The restricted vacuum expectation values $\langle  X \rangle_n^{^{\!(\mathds{1})}\!}(\lambda)$ 
defined in (\ref{rest_vev_free}) normalized to the form $V_n^{^{\,(\mathds{1})}\!}(\lambda)$ 
defined in (\ref{Vdef}) as a function of $\lambda$. The symbols represent the results for different intervals $I_n$ 
and the curves show the fits with $h(\Delta_n[\lambda - k_n^{^{\!(\mathds{1})}}])$. The data are for $V = 16 \times 16$,
$m = 0.1$ and $\mu = 0.05$ with an interval length of $\Delta_n = 1$.} 
\label{rest_vevs_plot}
\end{figure}

Having identified non-vanishing windows of $\lambda$ where we can safely evaluate the restricted vacuum expectation
values $\langle  X \rangle_n^{^{\!(\mathds{1})}\!}(\lambda)$ defined in (\ref{rest_vev_free}), we show some of these 
results for illustration in Fig.~\ref{rest_vevs_plot}. We plot the restricted vacuum expectations 
$\langle  X \rangle_n^{^{\!(\mathds{1})}\!}(\lambda)$ normalized to the form $V_n^{^{\,(\mathds{1})}\!}(\lambda)$ 
defined in (\ref{Vdef}) as a function of $\lambda$. The symbols represent the data that we determined in a small 
Monte Carlo simulation on a $16 \times 16$ lattice using $m = 0.1$ and $\mu = 0.05$. For $x$ we use intervals 
of length $\Delta_n = 1 \, \forall n$, such that the intervals are given by $I_n = [n, n+1]$. The symbols shown in  
Fig.~\ref{rest_vevs_plot} are the data for the intervals $I_n$ with $n = 0$, $n = 10$, $n = 20$, $n = 50$, $n = 80$ and 
$n = 120$. The 
lines are the fits of $V_n^{^{\,(\mathds{1})}\!}(\lambda)$ with $h(\Delta_n[\lambda - k_n^{^{\!(\mathds{1})}}])$
where $h(s)$ is defined in (\ref{hdef}).

\begin{figure}[t]
\centering
\hspace*{-2mm}
\includegraphics[width=85mm,clip]{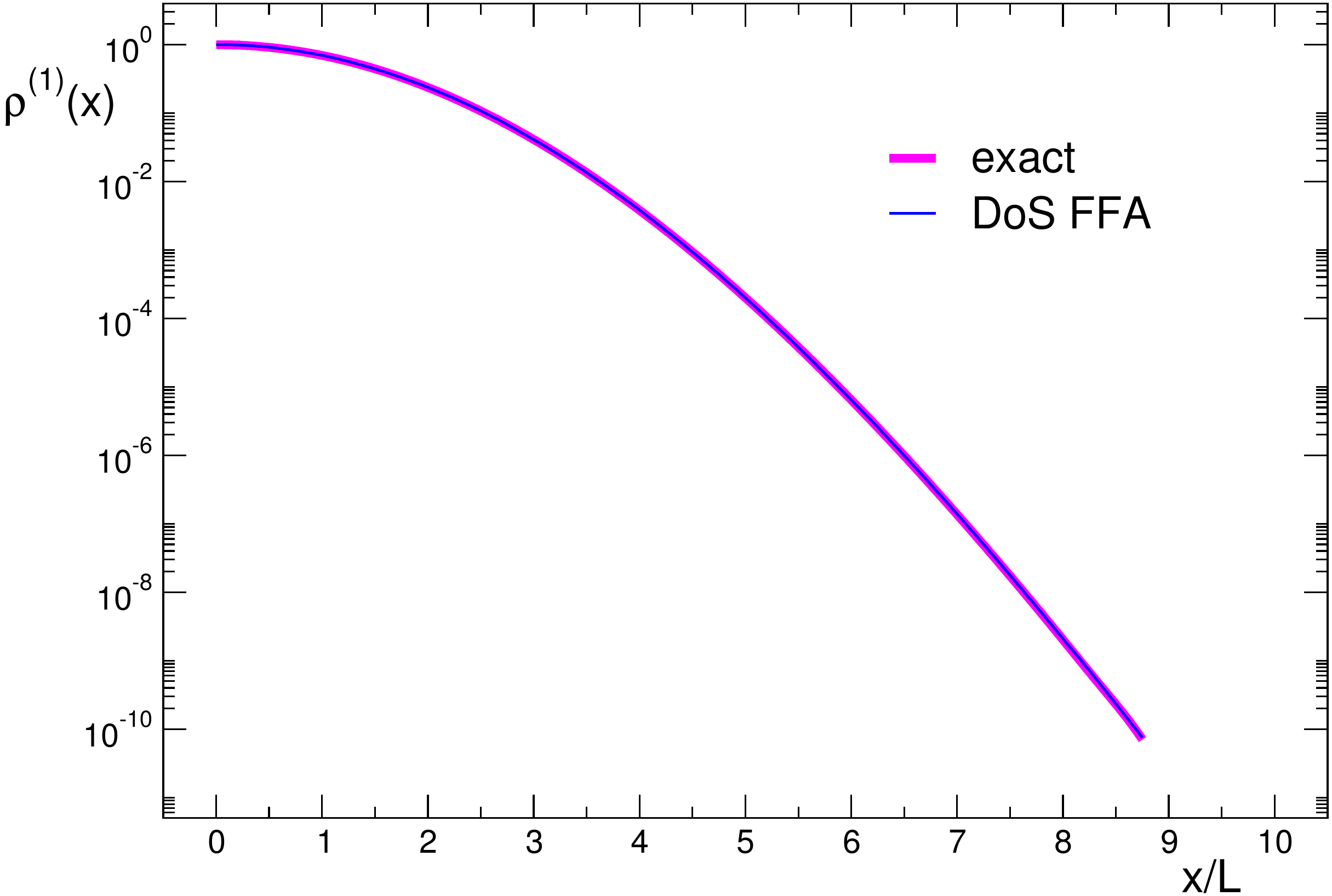}
\caption{$\rho^{^{(\mathds{1})}\!}(x)$ as a function of $x/L$, where $L = 16$ is the spatial extent of the lattice.
We compare the results from the DoS FFA calculation to the exact analytic solution. The data are for $V = 16 \times 16$,
$m = 0.1$ and $\mu = 0.05$. Note that we use a logarithmic scale for the vertical axis.} 
\label{lnrho_plot}
\end{figure}

From the fits of the restricted vacuum expectation value data with $h(\Delta_n[\lambda - k_n^{^{\!(\mathds{1})}}])$ 
we can determine all slopes $k_n^{^{(\mathds{1})}}$, and from those compute the density 
$\rho^{^{(\mathds{1})}\!}(x)$ using the closed expressions (\ref{rho_interval}). In Fig.~\ref{lnrho_plot} we show our
results for $\ln \rho^{^{(\mathds{1})}\!}(x)$ as a function of $x$, again using the DoS FFA data for  
$V = 16 \times 16$, $m = 0.1$ and $\mu = 0.05$. Note that this now is a quantity where for the free case we can 
compute analytical reference results. These are also shown in Fig.~\ref{lnrho_plot} and we find excellent agreement 
between the DoS FFA data and the analytic results, and stress at this point that we use a logarithmic scale on the 
vertical axis in Fig.~\ref{lnrho_plot}. 

We point out that further smoothening of the density with suitable fits will be part of a final strategy for DoS techniques --
see, e.g., the recent systematic comparison of such techniques in \cite{Francesconi:2019nph}.  

We conclude this section with commenting on how the analytic reference results shown in Fig.~\ref{lnrho_plot} 
were obtained: Starting from the definition (\ref{densities_free})
of the density $\rho^{^{(\mathds{1})\!}}(x)$ we may use the integral representation of the Dirac delta and find
\begin{eqnarray}
\rho{^{(\mathds{1})\!}}(x)  & \!\! = \!\! & \!\!\int \!\!{D}[\Phi] \; 
e^{\, - \, S_R[\Phi]} \; \delta \Big( x  -  \,X[\Phi] \Big) 
\nonumber  \\
& = & \int\limits_{-\infty}^\infty \!\! \frac{dq}{2\pi} \int \!\!{D}[\Phi] \; 
e^{\, - \, S_R[\Phi]} \; e^{\, - \, i q (x  -  \,X[\Phi] )} 
\nonumber \\
& = & 
\int\limits_{-\infty}^\infty \!\! \frac{dq}{2\pi} e^{ - i q x} \!\!  \int \!\!{D}[\Phi]  
e^{ - S_R[\Phi]  +  i q X[\Phi]} 
\nonumber \\
& = & 
\int\limits_{-\infty}^\infty \!\! \frac{dq}{2\pi} \; e^{ - i q x} \!\! \int \!\!{D}[\Phi] 
e^{ - \Phi^\dagger [ {\cal A} - i q {\cal B} ] \Phi} 
\nonumber \\
& \propto& 
\int\limits_{-\infty}^\infty \!\!\! dq \, \frac{e^{ \, - \, i q x}}{\det[{\cal A} - i q {\cal B} ]} \, ,
\label{densities_exact}
\end{eqnarray}
where ${\cal A}$, ${\cal B}$ are obtained from the matrices ${\cal A}[U,\mu]$ and ${\cal B}[U,\mu]$
defined in (\ref{ABelements}) by setting all links to $U_{\nu}(x) = \mathds{1}$. The determinant    
$\det[{\cal A} - i q {\cal B} ]$ can be computed with Fourier transformation and according to the 
last expression in (\ref{densities_exact}) the density $\rho{^{(\mathds{1})\!}}(x)$ is then obtained
as the Fourier transform of $1/\det[{\cal A} - i q {\cal B} ]$.

\section{DoS FFA for the canonical formulation of lattice QCD}

In this section we present the second new DoS approach to finite density lattice QCD, now working with the canonical 
ensemble. The canonical partition sums at different net-quark numbers are expressed as Fourier moments of the grand
canonical partition sum at imaginary chemical potential $\mu = i \theta /\beta$ and then $\theta$ is considered 
as an additional  degree of freedom in the path integral. In this form we may implement the DoS FFA and compute the
density $\rho(\theta)$.

\subsection{Canonical ensemble and density of states} 

The setting is as in the previous section, i.e., we study lattice QCD in $d$ dimensions with $N_f$ degenerate flavors 
of quarks, and the grand canonical partition sum $Z(\mu)$ is defined in (\ref{Z_mu}) -- (\ref{Hopdef}).
The canonical partition sums $Z_N$ at a fixed net quark number $N$ can be obtained as Fourier integrals over an
imaginary chemical potential $\mu = \I \, \theta^\prime / \beta$, where $\beta$ is the inverse temperature in lattice
units, i.e., $\beta = N_d$, with $N_d$ being the number of lattice points in time direction ($= d$-direction),
\begin{eqnarray}
Z_N & \! = \! & \! \int\limits_{-\pi}^\pi \! \frac{d\theta^\prime}{2\pi} \; Z(\mu) \, \bigg|_{\mu\,=\,\I\frac{\theta^\prime}{\beta}} \; 
e^{\, -i \theta^\prime  N} 
\label{ZN}  \\
&\! = \! & \! \int\limits_{-\pi}^\pi \! \frac{d\theta^\prime}{2\pi} \int \!\!  D[U]\,
e^{-S_g[U]} \, \det {\cal D}[U,\mu]^{\,N_f} \bigg|_{\mu\,=\,\I\frac{\theta^\prime}{\beta}} \; e^{\,-i \theta^\prime N} .
\nonumber
\end{eqnarray}
The corresponding  free energy density at fixed $N$ is defined as $f_N = - \ln Z_N / V$, $V = N_s^{d-1} \, N_d$.
Simple bulk observables can be obtained as derivatives of $f_N$ with respect to couplings of the theory. 
An example is the vacuum expectation value of the scalar fermion bilinear, 
$\langle \, \overline{\psi}(x) \psi(x) \, \rangle_N = \partial f_N / \partial m$, 
\begin{eqnarray}
\hspace*{-7mm}
&& \langle \, \overline{\psi}(x) \psi(x) \, \rangle_N 
  = \; - \frac{N_f}{V} 
\frac{1}{Z_N} \! \int\limits_{-\pi}^\pi \! \frac{d\theta^\prime}{2\pi} \int \! \! D[U] \;
e^{\, - \, S_g[U]}\; 
 \label{scalar} \\
\hspace*{-7mm} && \hspace{14mm} \times \; 
\det D[U,\mu]^{\, N_f} \; \mbox{Tr} \, {\cal D}^{-1}[U,\mu] \, \bigg|_{\mu\,=\,i \frac{\theta^\prime}{\beta}} \,
e^{\, - \, i \, \theta^\prime N} \; .
 \nonumber 
\end{eqnarray}
The derivative generates the insertion of Tr$\, D^{-1}[U,\mu]$, i.e., the traced inverse Dirac operator 
(quark propagator) as an additional factor in the path integral. Note that also in the quark propagator 
the chemical potential $\mu$ appears and is set to the complex value $\mu = i \theta^\prime/\beta$, 
used for projecting to fixed net quark number $N$. General vacuum expectation values at fixed $N$ 
have the form
\begin{eqnarray}
\hspace*{-7mm} && \langle  {\cal  O } \rangle_N \; = \; \frac{1}{Z_N} 
\int\limits_{-\pi}^\pi \! \frac{d\theta^\prime}{2\pi} \int \! \! D[U] \; e^{-S_g[U]} \;
\\
\hspace*{-7mm} && \hspace{14mm} \times \;
 \det {\cal D}[U,\mu]^{ \, N_f} \;  {\cal O} [U,\mu] \,  \bigg|_{\mu\,=\,i \frac{\theta^\prime}{\beta}} \; 
e^{- i \theta^\prime N} \;.
\nonumber
\end{eqnarray}
The expressions for the observables at fixed net quark number $N$ can be rewritten with the help of 
densities $\rho^{^{(J)}}\!(\theta)$ defined as (again normalization is ignored here)
\begin{equation}
\rho^{^{({\cal J})}\!}(\theta) 
\; =  \int \! \! D[U] \;
e^{-S_g[U]} \, \det {\cal D}[U,\mu]^{\, N_f} {\cal J}[U,\mu]\, 
\bigg|_{\mu\, = \, i \frac{\theta}{\beta}} .
\label{rho_candos}
\end{equation}
${\cal J}[U,\mu]$ is an arbitrary functional of the gauge fields, which, if it contains the quark propagator,
may also depend on the chemical potential $\mu$. Note that again different choices of ${\cal J}[U,\mu]$ result in 
different densities $\rho^{^{({\cal J})\!}}(\theta)$ and as before we use a superscript ${\cal J}$ to make clear which 
density we refer to. 

With the densities $\rho^{^{({\cal J})}\!}(\theta)$ vacuum expectation values $\langle {\cal O} \rangle_N$ at fixed net quark 
number can be expressed as
\begin{eqnarray}
\langle  {\cal  O} \rangle_N & = & \frac{1}{Z_N} \int\limits_{-\pi}^\pi \! d\theta \; \rho^{^{( {\cal O})}\!}(\theta)\,
e^{\, - i \theta N} \; , 
\nonumber \\
Z_N & = & \int \limits_{-\pi}^\pi  \! d\theta \; \rho^{^{(\mathds{1})}\!}(\theta)\,
e^{\, -i \theta N}\;.
\label{obsdens_new}
\end{eqnarray}
It is important to note that the densities $\rho^{^{({\cal J})}\!}(\theta)$ have symmetries that should be identified, 
because this allows one to reduce the range of $\theta$ that one needs to integrate over. Thus also the 
$\rho^{^{({\cal J})}}\!(\theta)$ need to be determined only in the reduced range of $\theta$ which lowers 
the numerical cost. In the previous section we have used charge conjugation symmetry to show that the density 
$\rho^{^{({\mathds{1}})}}\!(x)$ for the imaginary part $x \equiv X[U,\Phi]$ is even in $x$. Also here it is straightforward 
to establish that $\rho^{^{({\mathds{1}})}}\!(\theta)$ is even. As before we invoke the transformation property 
$C^{-1} {\cal D}[U^\prime\!,\mu] C = {\cal D}[U,-\mu]^T$  of the Dirac operator ${\cal D}[U,\mu]$ under charge conjugation,
where $C$ is the charge conjugation matrix and $U^\prime$ denotes the charge conjugate gauge links defined 
in (\ref{trafo}). Thus we find $\det {\cal D}[U^\prime\!,\mu] = \det {\cal D}[U,-\mu]$. Using the invariance 
$S_g[U^\prime] = S_g[U]$ and $\int D[U^\prime] = \int D[U]$  of gauge action and measure we conclude
\begin{eqnarray}
&& \rho^{^{({\mathds{1}})}\!}(-\theta) \, =  \! \int \! \! D[U^\prime] \,
e^{-S_g[U^\prime]} \, \det {\cal D}[U^\prime,\mu]^{\, N_f}  
\bigg|_{\mu\, = \, - i \frac{\theta}{\beta}} 
\nonumber \\
&& \hspace*{-7mm}
= \! \int \! \! D[U] \, e^{-S_g[U]} \, \det {\cal D}[U,-\mu]^{\, N_f}  
\bigg|_{\mu\, = \, - i \frac{\theta}{\beta}}  = \; \rho^{^{({\mathds{1}})}\!}(\theta) .
\label{rho_candos_sym}
\end{eqnarray}
In a similar way as in (\ref{rho_candos_sym}) one can show that also the general densities $\rho^{^{({\cal J})}\!}(\theta)$
are either even or odd functions, depending on the symmetry of the insertion ${\cal J}[U,\mu]$ (after decomposition into
$C$-even and $C$-odd parts if necessary). Thus the integrals (\ref{obsdens_new}) for evaluating observables only 
run from $0$ to $\pi$ and exploring charge conjugation symmetry cuts the numerical cost in half.

We conclude this subsection with discussing another interesting symmetry property of the density, 
which not necessarily can be used to reduce the numerical cost, but reflects an important aspect of
the underlying physics:  If QCD is in a purely
hadronic phase this is equivalent to 
$\rho^{^{(\mathds{1})}}\!(\theta)$ being $2\pi/3$ periodic. 
This property corresponds to the Roberge-Weiss symmetry and can be seen as follows: 
The statement that QCD is in a purely hadronic phase means that $Z_N = 0$ for all net quark numbers $N$ that 
are not multiples of $3$. We first assume that 
$\rho^{^{(\mathds{1})}}\!(\theta)$ is $2\pi/3$-periodic. Then we find
\begin{eqnarray}
Z_N & \!\!  =  \!\!  & \int \limits_{-\pi}^\pi \!\! \frac{d \theta}{2\pi} \, \rho^{^{(\mathds{1})}}\!(\theta) \, e^{-i \theta N} \; = 
\sum_{j=-1}^1  
\int \limits_{(2j-1)\pi/3}^{(2j+1)\pi/3} \!\! \frac{d \theta}{2\pi} \, \rho^{^{(\mathds{1})}}\!(\theta) \, e^{-i \theta N} 
\nonumber \\
&  \!\!  =  \!\!  & 
\sum_{j=-1}^1
\int \limits_{-\pi/3}^{\pi/3} \!\!\! \frac{d \theta}{2\pi} \rho^{^{(\mathds{1})}}\!\Big(\theta+\frac{2j\pi}{3}\Big) 
e^{-i \big(\theta + \frac{2j\pi}{3}\big)N} 
\nonumber \\
&  \!\!  =  \!\!  & 
\sum_{j=-1}^1 e^{ \, -i \frac{2j\pi}{3}\, N}
\int \limits_{-\pi/3}^{\pi/3} \!\! \frac{d \theta}{2\pi} \, \rho^{^{(\mathds{1})}}\!(\theta) \, e^{-i \theta N} 
\nonumber \\
&  \!\!  =  \!\!  &
\delta_{\;N\!\!\!\!\!\!\mod \!\!\,3  \,,\,   0} \; \;\;
3 \!\!\!\!\int \limits_{-\pi/3}^{\pi/3} \!\! \frac{d \theta}{2\pi} \, \rho^{^{(\mathds{1})}}\!(\theta) \, e^{-i \theta N} \; ,
\end{eqnarray}
which shows that a $2\pi/3$-periodic density $\rho^{^{(\mathds{1})}}\!(\theta)$ implies that only $Z_N$ where 
$N$ is a multiple of 3 are non-vanishing. 

For the inverse statement we can use the completeness and orthogonality 
of the Fourier factors $e^{i \theta N}$ and sum over $N$ the 
$Z_N$ in the form (\ref{obsdens_new}) with factors $e^{i \theta N}$ and find
\begin{equation}
\rho(\theta) \, = \, \sum_{N \in \mathds{Z}} Z_N \, e^{i \theta N} \, = \; Z_0 + \sum_{N=1}^\infty Z_N \, 2 \cos(\theta N) \; ,
\label{rho_ZNsum}
\end{equation}
where in the second step we used that $\rho^{^{(\mathds{1})}}\!(\theta)$ is even which in turn leads to $Z_N = Z_{-N}$.
The relation (\ref{rho_ZNsum}) implies that if the $Z_N$ vanish for values of $N$ which are not multiples of 3 the
density $\rho^{^{(\mathds{1})}}\!(\theta)$ is $2\pi/3$-periodic.

We remark, that the representation (\ref{rho_ZNsum}) of course holds in both, the hadronic and a possible non-hadronic 
phase, and in our small numerical test below we will use the form (\ref{rho_ZNsum}) to determine the canonical
partition sums $Z_N$ from a fit of the density according to  (\ref{rho_ZNsum}).

\subsection{Implementation of DoS FFA}

Having discussed the densities $\rho^{^{({\cal J})}}\!(\theta)$ and their symmetries we can now start the implementation 
of DoS FFA. For convenience we introduce the notation 
$\det {\cal D}[U,\theta^\prime] \equiv \det {\cal D}[U,\mu] \, |_{\mu\,=\,i \theta^\prime/\beta}$. 
For imaginary chemical potential 
$\gamma_5$-hermiticity guarantees that $\det {\cal D}[U,\theta^\prime]$ is real, such that the factor 
$\det {\cal D}[U,\theta^\prime]^{\,N_f}$ is real and positive for even $N_f$ (or sufficiently large mass in case $N_f$ is odd), 
and we may write 
\begin{equation}
e^{-S_g[U]}  \, \det {\cal D}[U,\theta^\prime]^{\,N_f} \; = \; e^{ \, - \, S_R[U,\theta^\prime]} \; ,
\label{SRcan}
\end{equation} 
with $S_R[U,\theta^\prime] \, \equiv \, S_g[U] \, - \, N_f \ln \det {\cal D}[U,\theta^\prime]$ such that $S_R[U,\theta^\prime]$ 
is real \footnote{In a practical implementation one will of course again use pseudo-fermion techniques to 
evaluate the determinant, but for notational convenience we here use the combined effective action 
$S_R[U,\theta^\prime]$.}. Using (\ref{SRcan}) we may write the canonical partition sum as 
\begin{equation}
Z_N \; = \; \int\limits_{-\pi}^\pi \frac{d\theta^\prime}{2\pi} \int \! \! D[U] \; 
e^{ \, - \, S_R[U,\theta^\prime] \, - \, i \, \theta^\prime N} \; .
\label{ZN2}
\end{equation}
It is interesting to note that the gauge fields $U_\nu(x)$ and the phase variable $\theta^\prime$ enter the 
path integral in the same way, i.e., both are integrated over in the path integral and appear in the exponent of the 
Boltzmann factor. Thus one may view $\theta^\prime$ as one more degree of freedom in the path integral and 
compare (\ref{ZN2}) with the generic form (\ref{Zbosonicdef}) used in the general discussion of the DoS FFA in Section 2. 
The exponent in the integral is the action $S[U,\theta^\prime]$ for all dofs.\ and we have already identified the 
real part of the action as $S_R[U,\theta^\prime]$. The imaginary part, which only depends on 
$\theta^\prime$, may be identified as  $X[\theta^\prime] = \theta^\prime$, and the parameter $\alpha$ in 
(\ref{Zbosonicdef}) is identified with the negative of the net quark number, i.e., $\alpha = - N$.

Having found a form of the problem that matches the generic form discussed in Section 2, we may identify the restricted 
vacuum expectation values needed for the determination of the densities $\rho^{^{({\cal J})\!}}(\theta)$. They
are given by
\begin{eqnarray}
\hspace*{-4mm} && \langle  \theta \rangle_n^{^{\!({\cal J})}\!}(\lambda)  =  
\frac{1}{Z_n^{^{\,({\cal J})}\!}(\lambda) }  \int\limits_{-\pi}^\pi \!\! d\theta^\prime \!\!
\int \!\! D[U] \, e^{ \, - \, S_R[U,\theta^\prime] \, + \, \lambda \, \theta^\prime} \, \theta^\prime \, \Theta_n(\theta^\prime) 
\nonumber \\
\hspace*{-4mm}  && =    
\frac{1}{Z_n^{^{\,({\cal J})}\!}(\lambda) } \! \int\limits_{\theta_n}^{\theta_{n+1}} \!\! d\theta^\prime \!\!
\int \!\! D[U] \;  e^{ \, - \, S_{g}[U] }  \; \det {\cal D}[U,\theta^\prime]^{N_f}  \; \theta^\prime \;
e^{\, \lambda \, \theta^\prime } ,
\nonumber \\
\hspace*{-5mm}  &&
\label{rest_vev_candos}
\end{eqnarray}
where, as mentioned before, the fermion determinant may be represented using pseudo-fermions. The restricted 
vacuum expectation values (\ref{rest_vev_candos}) do not have a complex action problem and may be computed with 
standard Monte Carlo techniques. Note that the imaginary chemical potential $\theta^\prime$ 
is an additional degree of freedom that is restricted to the interval $I_n = [\theta_n, \theta_{n+1}]$ and needs to be 
updated as well. 

After evaluating the restricted vacuum expectation values $\langle  \theta \rangle_n^{^{\!({\cal J})}\!}(\lambda)$ they 
need to be brought into the normalized form $V_n^{^{\,({\cal J})}\!}(\lambda)$ (in (\ref{Vdef}) replace 
$\langle  X \rangle_n^{^{\!({\cal J})}\!}(\lambda)$ by $\langle  \theta \rangle_n^{^{\!({\cal J})}\!}(\lambda)$
and $x_n$ by $\theta_n$), such that they can be fit with $h(\Delta_n[\lambda - k_n^{^{\!({\cal J})}}])$, which leads
to the slopes $k_n^{^{\!({\cal J})}}$. From the slopes the densities $\rho^{^{({\cal J})}}\!(\theta)$ are computed using 
$(\ref{piecewise})$, $(\ref{rho_interval})$, and finally observables in the canonical picture at fixed net quark number 
$N$ are obtained from the densities via (\ref{obsdens_new}).

\subsection{Tests of the canonical DoS FFA in the free case}

\begin{figure}[t]
\centering
\hspace*{-2mm}
\includegraphics[width=85mm,clip]{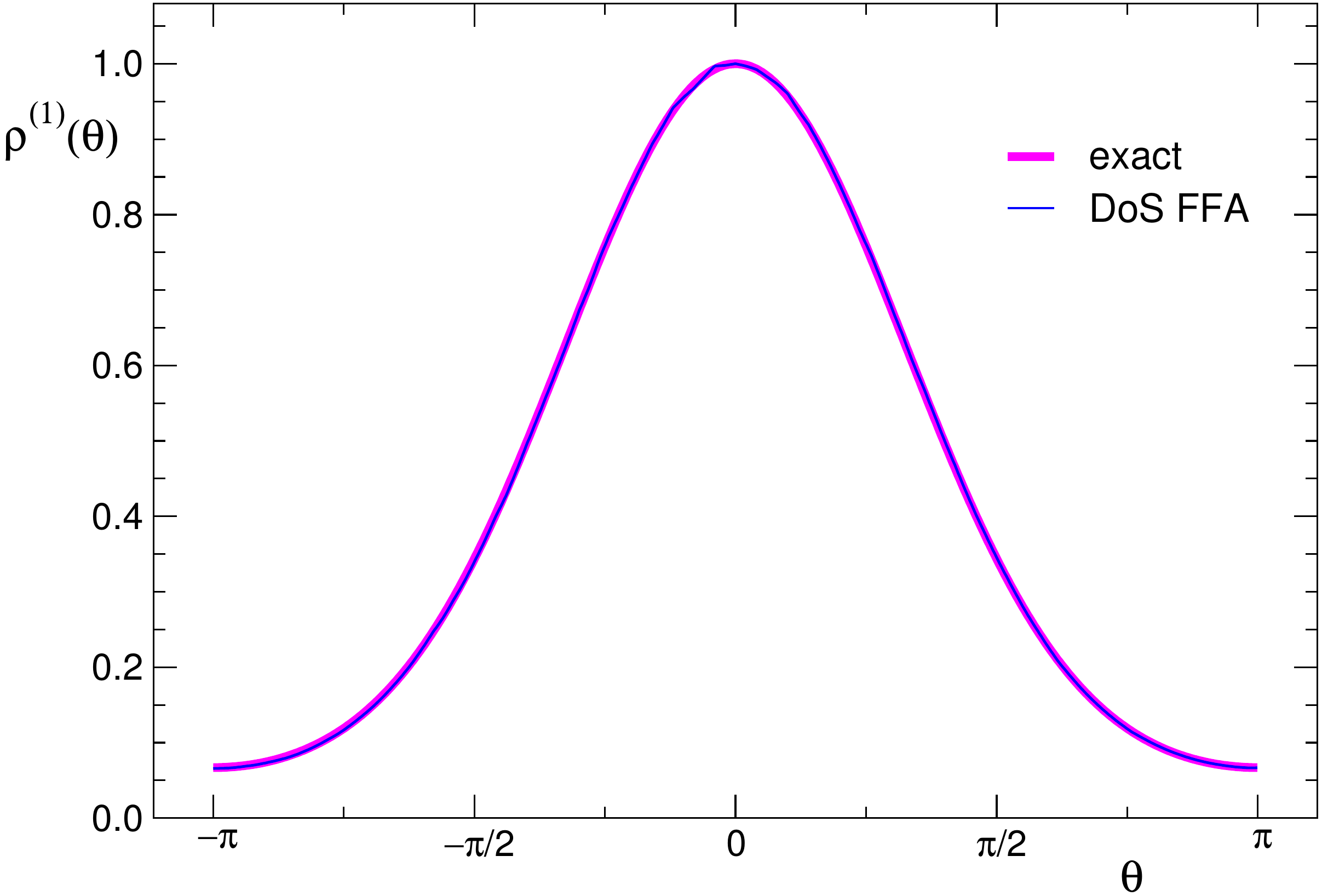}
\caption{The density $\rho^{^{({\mathds{1}})}}\!(\theta)$ as a function of $\theta$. We compare the DoS FFA result 
(thin blue curve) with the exact result (thick magenta curve). Note that we did not use the fact that the density 
is known to be an even function and for evaluation purposes numerically determined 
$\rho^{^{({\mathds{1}})}}\!(\theta)$  in the full range $\theta \in [-\pi,\pi]$.} 
\label{rho_CanDoS}
\end{figure}

Again we use 2-d free fermions at finite density for a first test also in the canonical formulation of the DoS FFA.
For the free case the density (\ref{rho_candos}) for the choice ${\cal J} = \mathds{1}$ reduces to the particularly
simple expression (we set $N_f = 2$) 
\begin{equation}
\rho^{^{({\mathds{1}})}}\!(\theta) \; = \; \det {\cal D}[\mu]^{\,2} \, 
\bigg|_{\mu\, = \, i \frac{\theta}{\beta}} \; ,
\label{rho_candos_free}
\end{equation}
where ${\cal D}[\mu]$ denotes the Wilson Dirac operator (\ref{Diracdef}), (\ref{Hopdef}) in $d = 2$ with all link 
variables set to $U_\mu(x) = \mathds{1}$. It is straightforward to evaluate this quantity using Fourier transformation and the 
reference data used in Fig.~\ref{rho_CanDoS} below for verification were computed in this way.

\begin{figure}[t]
\centering
\hspace*{-2mm}
\includegraphics[width=85mm,clip]{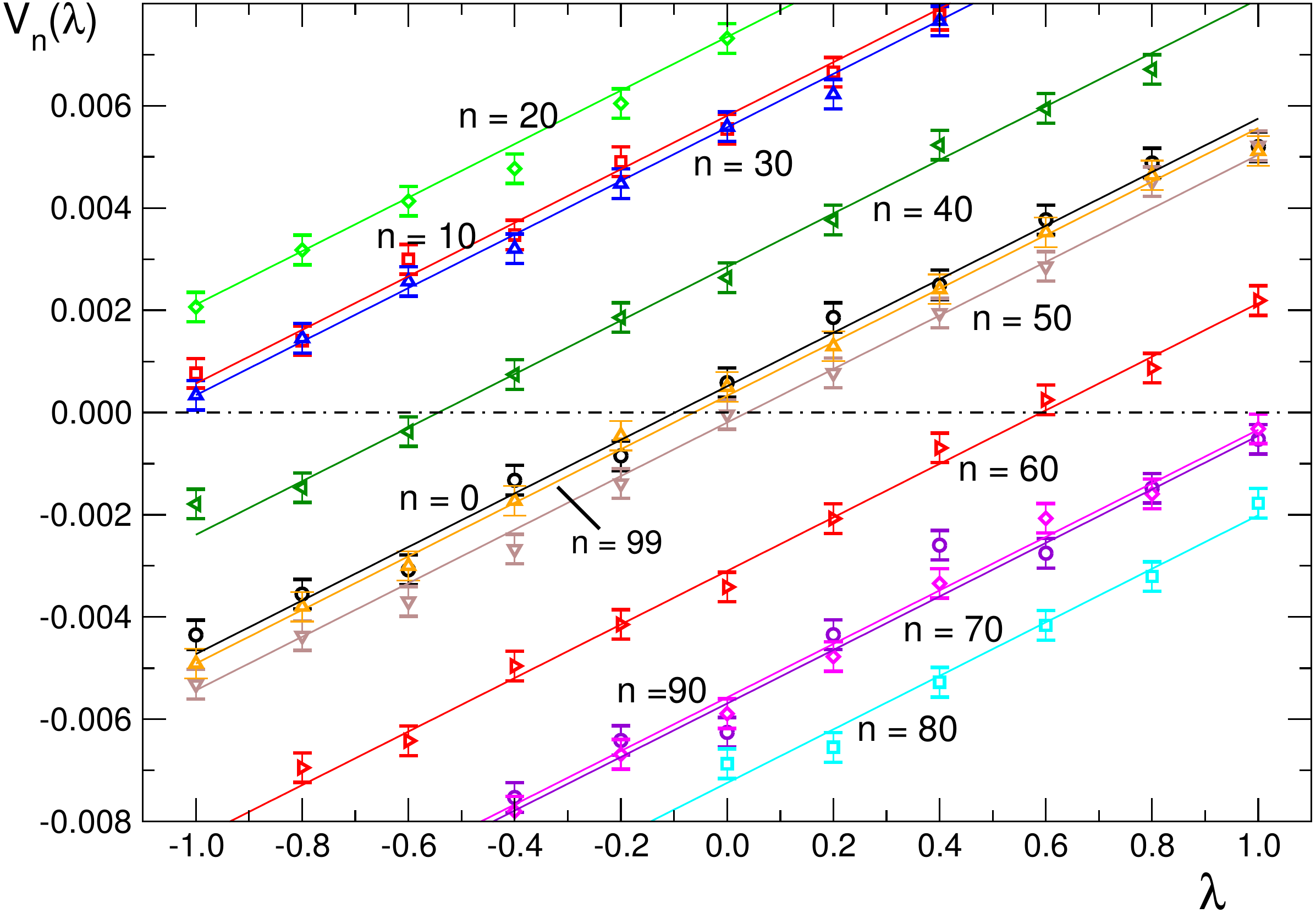}
\caption{The restricted vacuum expectation values $\langle \theta  \rangle_n^{^{\!(\mathds{1})}\!}(\lambda)$ 
defined in (\ref{rest_vev_candos}) normalized to the form $V_n^{^{\,(\mathds{1})}\!}(\lambda)$ 
defined in (\ref{Vdef}), plotted as a function of $\lambda$. The symbols represent the results for different intervals 
$I_n \subset [-\pi,\pi]$ 
(see the labels next to the data) and the curves the fits with $h(\Delta_n[\lambda - k_n^{^{\!(\mathds{1})}}])$. 
The data are for $V = 16 \times 16$, $m = 0.1$ and $\mu = 0.05$ with an interval length of $\Delta_n = 2\pi/100$.} 
\label{Vn_CanDoS}
\end{figure}

The restricted vacuum expectation values $\langle  \theta \rangle_n^{^{\!({\mathds{1}})}\!}(\lambda)$ defined in 
(\ref{rest_vev_candos}) reduce to 
\begin{equation}
\langle  \theta \rangle_n^{^{\!({\mathds{1}})}\!}(\lambda) \; = \;  
\frac{1}{Z_n^{^{({\mathds{1}})}}\!(\lambda) } \! \int\limits_{\theta_n}^{\theta_{n+1}} \! d\theta^\prime
\; \det {\cal D}[U,\theta^\prime]^{2}  \; \theta^\prime \;
e^{\, \lambda \, \theta^\prime } \; .
\label{rest_vev_candos_free}
\end{equation}
For a test of the canonical DoS FFA formulation we evaluated the restricted expectation values in a small 
Monte Carlo simulation on $16 \times 16$ lattices with mass $m = 0.1$. Although we could use the 
symmetry of the density $\rho^{^{({\mathds{1}})}}\!(\theta)$ and restrict the determination of 
$\rho^{^{({\mathds{1}})}}\!(\theta)$ to the interval $\theta \in [0,\pi]$, we here determine the density for  
the full range $\theta \in [-\pi,\pi]$. The symmetry of  $\rho^{^{({\mathds{1}})}}\!(\theta)$ should emerge 
and serves as a consistency check for the calculation. The interval $[-\pi, \pi]$ was divided into 
100 equal size intervals $I_n$ of length $\Delta_n = 2\pi/100 \; \forall n$.  The Monte Carlo simulation 
for sampling the restricted $\theta$-integral in each interval $I_n$ uses a statistics 
of $10^6$ sweeps of local Metropolis updates separated by 20 sweeps for decorelation and $10^5$ sweeps for 
initial equilibration. The determinant in the acceptance step was computed with Fourier transformation 
und we typically use 10 values of $\lambda$ for the evaluation of the restricted vacuum 
expectation values $\langle  \theta \rangle_n^{^{\!({\mathds{1}})}\!}(\lambda)$.

In Fig.~\ref{Vn_CanDoS} we show the results for the restricted vacuum expectation values 
$\langle  \theta \rangle_n^{^{\!({\mathds{1}})}\!}(\lambda)$ already in their normalized form 
$V_n^{^{\!({\mathds{1}})}\!}(\lambda)$  according to (\ref{Vdef}). The symbols represent the data from
the Monte Carlo simulation and the full curves are the fits with $h(\Delta_n[\lambda - k_n^{^{\!({\mathds{1}})}}])$
according to (\ref{hdef}). The values of $\lambda$ where the curves cross 0 are the slopes 
$k_n^{^{\!({\mathds{1}})}}$. These crossing points start near 0 for the smallest $n$ (i.e., intervals $I_n$ near $-\pi$) 
become negative then, revert back to 0, move to positive values and finally revert again back to $0$ for 
intervals $I_n$ near $+\pi$. This full oscillation of the corresponding slopes $k_n^{^{\!({\mathds{1}})}}$ reflects 
the $2\pi$-periodicity of the density $\rho^{^{({\mathds{1}})}}\!(\theta)$ (compare Fig.~\ref{rho_CanDoS}).

From the slopes $k_n^{^{\!({\mathds{1}})}}$ obtained with the fits of the restricted vacuum expectation values we 
determined the density $\rho^{^{({\mathds{1}})}}\!(\theta)$ using $(\ref{piecewise})$ and $(\ref{rho_interval})$. 
In Fig.~\ref{rho_CanDoS} we compare the density determined in this way with the analytic result from 
Fourier transformation. The analytic result is represented by the thick magenta curve on top of which 
we plot the DoS result (thin blue curve). We stress again that the density $\rho^{^{({\mathds{1}})}}\!(\theta)$  
was determined with DoS FFA for the full range $\theta \in [-\pi, \pi]$ and the fact that $\rho^{^{({\mathds{1}})}}\!(\theta)$ 
indeed comes out as an even function is a consistency check of the method. Anyway, the much more stringent test is the
comparison with the analytic result where the plot shows that the DoS FFA curve perfectly falls on top of the exact curve
determined as discussed above.

We complete our first test of the canonical DoS formulation with FFA by evaluating the canonical partition sums $Z_N$ 
from the density $\rho^{^{({\mathds{1}})}}\!(\theta)$ via the integrals (\ref{obsdens_new}) and comparing these Monte Carlo 
based results to the exact calculation based on a direct evaluation of (\ref{ZN}) with Fourier transformation techniques.  

\begin{figure}[t]
\centering
\hspace*{-2mm}
\includegraphics[width=85mm,clip]{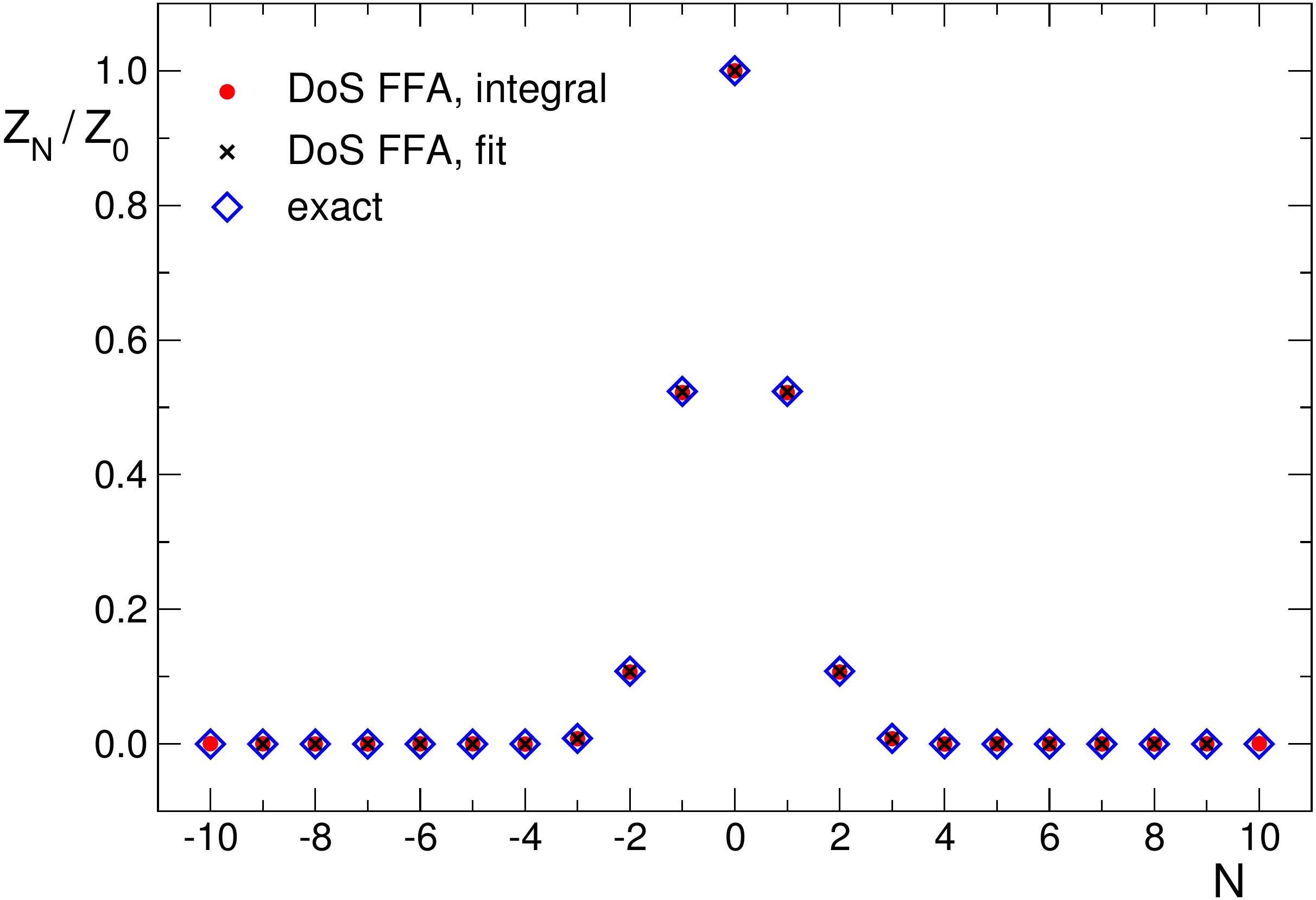}
\caption{Distribution of the canonical partition sums $Z_N$. We normalize the $Z_N$ by $Z_0$ and plot them as a 
function of $N$. The blue diamonds are the reference data from an exact evaluation and we show the results from 
two types of DoS FFA determinations: The red circles are from the integrals  (\ref{obsdens_new}), while the black 
crosses were determined from a fit of $\rho^{^{({\mathds{1}})}}\!(\theta)$ with the form (\ref{rho_ZNsum}).} 
\label{ZNplot}
\end{figure}

In Fig.~\ref{ZNplot} we show the corresponding results for $Z_N$ normalized with $Z_0$ as a function of $N$. 
The blue diamonds represent the exact results and the red dots the DoS FFA data obtained with the 
integrals (\ref{obsdens_new}). The distribution resembles a Gaussian, rapidly decreasing with increasing $|N|$ 
(which is of course a volume dependent statement). We find that the DoS FFA data based on 
(\ref{obsdens_new}) match the exact results very well. 

We have already pointed out in the discussion of the direct DoS FFA approach that fitting the density with a
suitable function will be an important part of future DoS strategies. Usually a large polynomial would be used 
for such a fit (see, e.g., \cite{Francesconi:2019nph,FFA_2,FFA_3} for related discussions), but for the canonical
DoS approach the representation (\ref{rho_ZNsum}) of the density suggests another option for a fit, namely using 
a superposition of cosines (sines for odd densities). For the particular case of the density  
$\rho^{^{({\mathds{1}})}}\!(\theta)$ the fit parameters are the canonical partition sum $Z_N$. In order to test this 
possibility, we determined the $Z_N$ also from a fit of $\rho^{^{({\mathds{1}})}}\!(\theta)$ with (\ref{rho_ZNsum}).
The corresponding results are shown as black circles in Fig.~\ref{ZNplot} and again we find a very good agreement 
with the analytical results.  This demonstrates that smoothening techniques based on periodic representations
of the type (\ref{rho_ZNsum}) should be an interesting option to be explored in future development of canonical DoS 
techniques. 

Also for the CanDos we would like to stress that the tests presented here constitute merely a very first 
assessment of the new approach and only the implementation in a full QCD simulation will show how well the 
numerical challenges can be brought under control in a calculation that includes the full gauge field dynamics.  

\section{Summary, discussion and outlook}

In this article we have discussed two proposals for a modern DoS approach to finite density lattice QCD based on 
representations of the theory with pseudo-fermions. In the direct grand canonical approach the fermion determinant 
is represented with pseudo-fermions and subsequently their effective action is separated into real and imaginary parts
such that the latter can then directly be treated with DoS FFA. We worked out the details of the formulation and 
provided some bounds on the involved kernels of the pseudo-fermion bilinears, showing that the method is applicable in 
an interesting range of values of the chemical potential $\mu$. We presented very preliminary tests in the free case where 
a comparison to exact results allows one to assess the new approach. The direct DoS formulation in the grand canonical 
picture is rather straightforward, but has the disadvantage that also the densities depend on the chemical potential 
$\mu$. As a consequence the densities have to be re-calcuated when changing $\mu$. Whether this approach can 
beat our second suggestion, the canonical version of DoS FFA, has to be seen in future more detailed tests.

In the canonical formulation observables at a fixed net quark number $N$ are obtained as the Fourier moments of the 
partition sum at imaginary chemical potential $\mu = i \theta /\beta$. In this setting we promote the angle $\theta$ 
to a new dynamical variable and interpret the exponent of the Fourier factors $e^{- i \theta N}$ as the imaginary part
of the action. Again we treat this imaginary part with the DoS FFA approach and compute the density $\rho(\theta)$ 
as a function of $\theta$. Observables at different net particle numbers $N$ are then obtained by integrating the same 
density $\rho(\theta)$ with different Fourier factors $e^{- i \theta N}$. Obviously here the resulting density $\rho(\theta)$ 
can be used for different net particle numbers $N$, but of course the accuracy of the determination of $\rho(\theta)$ 
has to be higher for larger $N$. Also here further tests that go beyond the first numerical checks we have presented 
here will be necessary to assess whether this formulation will be able to compete with other approaches to finite 
density QCD. 

Both formulations we have suggested here, for the first time implement DoS techniques directly in a pseudo-fermion
representation. This has the advantage that these well established techniques can be used in the framework of a modern 
DoS setting (here the DoS FFA is used but it is also straightforward to implement the ideas proposed here in the LLR 
framework). Obviously the simple exploratory numerical tests we have presented in this paper only serve to check the 
plausibility of the 
two new formulations and a much more detailed assessment will be necessary to explore their potential. Such further 
numerical tests are currently in preparation. 

We conclude with remarking that the techniques developed here go beyond applications to finite density QCD. The 
two approaches are general and can be applied to any lattice field theory with fermions where the interaction can be 
written with the help of a bosonic field such that the fermion action has a bilinear form and a fermion determinant
emerges when integrating out the fermions. The bosonic fields do not have to be gauge fields, but also auxiliary fields
of a Hubbard-Stratonovich transformation of quartic fermion interactions are a suitable option. 
We have begun to explore also these possible applications of the newly proposed DoS FFA techniques. 
Finally we remark that very recently \cite{Gattringer:2019egx} we
presented a first test of the new approaches, now for the case of lattice QCD formulated with staggered fermions.

\vskip5mm
\noindent
{\bf Acknowledgments:} We thank Mario Giuliani, Kurt Langfeld and Biagio Lucini for interesting discussions on the 
subject of DoS techniques. This work is supported by the Austrian Science Fund FWF, grant I 2886-N27 and 
partly also by the FWF DK 1203 ''Hadrons in Vacuum Nuclei and stars''.


\begin{thebibliography}{12}

\bibitem{Gocksch:1987nt}
  A.~Gocksch, P.~Rossi and U.M.~Heller,
  Phys.\ Lett.\ B {\bf 205} (1988) 334.

\bibitem{Gocksch:1988iz}
  A.~Gocksch,
  Phys.\ Rev.\ Lett.\  {\bf 61} (1988) 2054.

\bibitem{Schmidt:2005ap}
  C.~Schmidt, Z.~Fodor and S.D.~Katz,
  PoS LAT {\bf 2005} (2006) 163
  [hep-lat/0510087].

\bibitem{Fodor:2007vv}
  Z.~Fodor, S.D.~Katz and C.~Schmidt,
  JHEP {\bf 0703} (2007) 121   
  [hep-lat/0701022].

\bibitem{Ejiri:2007ga}
  S.~Ejiri,
  Phys.\ Rev.\ D {\bf 77} (2008) 014508
  [arXiv:0706.3549].

\bibitem{Ejiri:2012ng}
  S.~Ejiri {\it et al.} [WHOT-QCD Collaboration],
  Central Eur.\ J.\ Phys.\  {\bf 10} (2012) 1322
  [arXiv:1203.3793].

\bibitem{WangLandau}
  F.~Wang and D.P.~Landau,
  Phys.\ Rev.\ Lett.\  {\bf 86} (2001)  2050
  [cond-mat/0011174].

\bibitem{Langfeld:2012ah}
  K.~Langfeld, B.~Lucini and A.~Rago,
  Phys.\ Rev.\ Lett.\  {\bf 109} (2012) 111601
  [arXiv:1204.3243].

\bibitem{Langfeld:2013xbf}
  K.~Langfeld and J.M.~Pawlowski,
  Phys.\ Rev.\ D {\bf 88} (2013) 071502
  [arXiv:1307.0455].
 
\bibitem{Langfeld:2014nta}
  K.~Langfeld and B.~Lucini,
  Phys.\ Rev.\ D {\bf 90} (2014) 094502
  [arXiv:1404.7187].

\bibitem{Langfeld:2015fua}
  K.~Langfeld, B.~Lucini, R.~Pellegrini and A.~Rago,
  Eur.\ Phys.\ J.\ C {\bf 76} (2016) 306
  [arXiv:1509.08391].

\bibitem{Garron:2016noc}
  N.~Garron and K.~Langfeld,
  Eur.\ Phys.\ J.\ C {\bf 76} (2016)  569
  [arXiv:1605.02709].

\bibitem{Garron:2017fta}
  N.~Garron and K.~Langfeld,
  Eur.\ Phys.\ J.\ C {\bf 77} (2017)  470
  [arXiv:1703.04649].
  
\bibitem{Francesconi:2019nph} 
  O.~Francesconi, M.~Holzmann, B.~Lucini and A.~Rago,
  arXiv:1910.11026 [hep-lat].

\bibitem{FFA_1}
  C.~Gattringer, M.~Giuliani, A.~Lehmann and P.~T\"orek,
  POS LATTICE {\bf 2015} (2016) 194
  [arXiv:1511.07176].

\bibitem{FFA_2}
  M.~Giuliani, C.~Gattringer and P.~T\"orek,
  Nucl.\ Phys.\ B {\bf 913} (2016) 627
  [arXiv:1607.07340].
  
  \bibitem{FFA_3}
  M.~Giuliani and C.~Gattringer,
  Phys.\ Lett.\ B {\bf 773} (2017) 166
  [arXiv:1703.03614].

\bibitem{Z3_FFA_1}
  Y.~Delgado Mercado, P.~T\"orek and C.~Gattringer,
  PoS LATTICE {\bf 2014} (2015) 203
  [arXiv:1410.1645].

\bibitem{Z3_FFA_2}
  C.~Gattringer and P.~T\"orek,
  Phys.\ Lett.\ B {\bf 747} (2015) 545
  [arXiv:1503.04947].

\bibitem{Gattringer:2016kco}
  C.~Gattringer and K.~Langfeld,
  Int.\ J.\ Mod.\ Phys.\ A {\bf 31} (2016)  1643007
  [arXiv:1603.09517].
  
\bibitem{Cheb1}
T.~ A.~Manteuffel, 
Num.\ Math.\ {\bf 28} (1977) 307.

\bibitem{Cheb2}
Y.~Saad, 
''Iterative Methods for Sparse Linear Systems'', 
Society for Industrial and Applied Mathematics, Philadelphia, PA, USA, 2nd ed., 2003.

\bibitem{Gattringer:2019egx} 
  C.~Gattringer, M.~Mandl and P.~T{\"o}rek,
  arXiv:1912.05040 [hep-lat].

\end{thebibliography}
\end{document}